\documentclass{aastex}
\usepackage{emulateapj5}
\usepackage{apjfonts}

\slugcomment{Accepted for publication in ApJ}

\newcommand{\pfrac}[2]{{\left(\frac{#1}{#2}\right)}}
\newcommand{\parens}[1]{{\left(#1\right)}}

\renewcommand{\min}{{\rm min}}
\renewcommand{\max}{{\rm max}}

\newcommand{\rmin}{{r_\min}}

\newcommand{\kT}{{k_B T_e}}
\newcommand{\MMd}{{M\!\dot{M}}}

\newcommand{\disk}{{\rm disk}}
\newcommand{\visc}{{\rm visc}}
\newcommand{\inc}{{\rm inc}}

\newcommand{\fcol}{{f_{\rm col}}}

\newcommand{\rt}{{\tilde{r}}}
\newcommand{\Mdot}{{\dot{M}}}
\newcommand{\Msun}{{M$_\odot$}}

\newcommand{\ASCA}{{\em ASCA}}
\newcommand{\EUVE}{{\em EUVE}}
\newcommand{\INTEGRAL}{{\em INTEGRAL}}
\newcommand{\Ginga}{{\em Ginga}}
\newcommand{\IUE}{{\em IUE}}
\newcommand{\Newton}{{\em XMM-Newton}}
\newcommand{\PCA}{{PCA}}
\newcommand{\RXTE}{{\em RXTE}}

\begin{document}

\title{Using Multiwavelength Observations to Determine the Black Hole
       Mass and Accretion Rate in the Type 1 Seyfert Galaxy NGC 5548}

\author{James Chiang\altaffilmark{1,2}}
\affil{NASA/GSFC, Code 661, Greenbelt MD 20771}

\author{Omer Blaes}
\affil{Department of Physics, University of California, 
       Santa Barbara CA 93106-9530}

\altaffiltext{1}{University of Maryland, Baltimore County, JCA/Physics Dept.
1000 Hilltop Circle, Baltimore MD 21250}
\altaffiltext{2}{Mailing Address: Stanford Linear Accelerator Center, 
MS 43A, 2575 Sand Hill Rd, Menlo Park CA 94025}

\begin{abstract}
We model the spectral energy distribution of the type 1 Seyfert galaxy
NGC 5548, fitting data from simultaneous optical, UV, and X-ray
monitoring observations.  We assume a geometry consisting of a hot
central Comptonizing region surrounded by a thin accretion disk.  The
properties of the disk and the hot central region are determined by
the feedback occurring between the hot Comptonizing region and thermal
reprocessing in the disk that, along with viscous dissipation,
provides the seed photons for the Comptonization process.  The
constraints imposed upon this model by the multiwavelength data allow
us to derive limits on the central black hole mass, $M \la 2 \times
10^7$~\Msun, the accretion rate, $\Mdot \la 2.5\times
10^5$~M$_\odot^2$yr$^{-1}/M$, and the radius of the transition region
between the thin outer disk and the geometrically thick, hot inner
region, $\sim 2$--5$\times 10^{14}$~cm.
\end{abstract}

\keywords{galaxies: active --- galaxies: individual (NGC 5548) 
          --- galaxies: Seyfert --- X-rays: galaxies}

\section{Introduction}

In two previous papers, we developed a model of thermal Comptonization
and disk thermal reprocessing to explain the data from recent
multiwavelength monitoring observations of the type 1 Seyfert galaxies
NGC~3516 and NGC~7469 (Chiang \& Blaes 2001, hereafter Paper~I; Chiang
2002, hereafter Paper~II).  The essential features of this model were
proposed by Poutanen, Krolik, \& Ryde (1997) to account for the
different spectral states of the Galactic X-ray binary Cygnus~X-1; and
a simplified version was used by Zdziarski, Lubi\'nski, \& Smith
(1999) to model the correlation seen in X-ray binaries and type 1
Seyfert galaxies between the X-ray spectral index and the relative
strength of the Compton reflection continuum.  A key element of this
model is the feedback occurring between a centrally located
Comptonizing plasma and the thin accretion disk that encircles it.
The disk reprocesses X-rays from the central region, thus providing a
significant fraction of the seed photons for the Comptonization
process itself.  For a given set of optical, UV, and X-ray data, this
feedback mechanism determines the equilibrium state of the
Comptonizing plasma.  By computing the equilibria required by the NGC
3516 and NGC 7469 data, we found that the seemingly uncorrelated
optical/UV and X-ray continuum variations exhibited by these sources
could be described almost entirely in the context of this model.
Furthermore, in modeling these variations, we were able to constrain
the central black hole masses and the accretion rates of these
objects; and we were able to compute the X-ray continuum at energies
extending beyond $\sim 20$~keV, a part of the spectrum that will soon
be made more accessible by the {\em International Gamma-Ray
Astrophysics Laboratory (INTEGRAL)}.

During an intensive three day monitoring campaign in 1998 April with
the {\em Rossi X-ray Timing Explorer (RXTE)}, the {\em Advanced
Satellite for Cosmology and Astrophysics (ASCA)}, and the {\em Hubble
Space Telescope}, NGC~3516 exhibited an optical continuum that was
surprisingly constant, consisting of only $\sim 3$\% variations, while
simultaneously showing $\sim 60$\% changes in the 2--10~keV X-ray flux
(Edelson et al.\ 2000).  This apparent lack of response by the optical
continuum to the changes in the X-ray flux has been taken as evidence
against the thermal reprocessing model for the optical and UV
variability.  However, despite the relatively small amplitude of its
fluctuations, the optical continuum varied essentially simultaneously
across a wide range of wavelengths, in a manner consistent with disk
thermal reprocessing (e.g., Krolik et al.\ 1991; Courvoisier \& Clavel
1991; Collin-Souffrin 1991).  In Paper~I, we showed that a constant
optical continuum in the presence of large X-ray fluctuations could be
reconciled with the disk thermal reprocessing model if the radial
extent, $r_s$, of the central Comptonizing plasma varied inversely
with the X-ray luminosity, $L_x$, so that the flux incident upon the
disk at moderately large radii remained constant.  Slight departures
from the $L_x \propto r_s^{-1}$ relationship would then produce the
highly correlated, low amplitude variations seen in the optical
continuum of NGC 3516.  We found that the range of X-ray fluxes and
spectral indexes exhibited by this source is naturally accommodated by
this behavior.

The interpretation of the NGC 7469 data is more complex.  Data from a
30 day monitoring campaign in 1996 June--July with the {\em
International Ultraviolet Explorer (IUE)} and \RXTE\ showed that the
UV continuum at $\lambda$1350\AA\ and the 2--10~keV X-ray light curves
were not correlated in any simple way (Nandra et al.\ 1998).  Although
the amplitude and time scales of the most prominent variations in the
two datasets were comparable, some features in the UV occurred nearly
simultaneously with similar looking features in the X-rays, while
other features that were similar in the two light curves were offset
by $\sim 4$~days.  A quantitative analysis revealed that the strongest
correlation between the two light curves corresponded to the UV
variations {\em leading} those of the X-rays by several days, contrary
to naive expectations of the thermal reprocessing model.  Nandra et
al.\ (2000) were able to explain these data, at least qualitatively,
by noting that the UV fluxes were highly correlated at near zero
temporal lag with the X-ray spectral index and also with a broad band
(0.1--100~keV) extrapolation of the 2--10~keV X-ray power-law.  When
we applied the thermal Comptonization/disk reprocessing model to these
data, we found that rather large changes in the bolometric X-ray
luminosity were required in order to produce the relatively large
variations seen in the UV (Paper~II).  Since the X-ray continua for
this source are relatively soft and are constrained in the 2--10~keV
band by the \RXTE/\PCA\ observations, a broad range of temperatures
for the Comptonizing electrons is needed so that the thermal
Comptonization process can produce a concomitantly large range of
X-ray luminosities.

Despite the difficulties associated with explaining these
observations, the application of the model to these data yielded some
encouraging results.  First, the relatively large inner truncation
radius of the thin disk, $\rmin \sim 3\times 10^{14}$~cm, that we
found by fitting the optical and UV continuum data for NGC 7469 is
consistent with that inferred from \ASCA\ observations of the narrow
iron K$\alpha$ emission line by Guainazzi et al.\ (1994) if the broad
line region reverberation mapping estimates of the black hole mass are
correct (Peterson \& Wandel 2000).  Second, the value we inferred for
the product of the black hole mass and accretion rate, $\MMd \sim
10^6$~M$_\odot^2$~yr$^{-1}$, agrees with that found by Collier et al.\
(1998) who examined time delays between UV and optical continuum bands
under the assumption of thermal reprocessing.  Third, a global energy
balance condition is satisfied by our models for both NGC~3516 and
NGC~7469 wherein the total bolometric luminosity is consistent with
the inferred accretion rate assuming nominal black hole masses of
$\sim 10^7$\Msun\ for each source.

In the present paper, we analyze simultaneous optical, UV, and X-ray
data for the Seyfert 1 galaxy NGC 5548 which were taken during two
separate monitoring observations: a ground-based-optical/\IUE/\Ginga\
campaign in 1989--90 (Peterson et al.\ 1991; Clavel et al.\ 1991,
1992; Nandra et al.\ 1991) and our own 1998 ground-based-optical/{\em
Extreme Ultraviolet Explorer (EUVE)}/\ASCA/\RXTE\ monitoring campaign
(Chiang et al.\ 2000; Dietrich et al.\ 2001).  Using our model, these
data allow us to place limits on the mass and accretion rate of this
object; and from that information, we can make predictions for what
should be seen by \Newton\ observations of this source in the optical,
UV, and X-rays and by possible forth-coming simultaneous observations
by \INTEGRAL\ with other X-ray satellites such as \Newton, the {\em
Chandra X-ray Observatory}, and \RXTE.  In \S2 of this paper, we
describe the optical, UV, and X-ray data.  In \S3, we present our
semi-analytic model of the spectral energy distributions (SEDs); and
in \S4, we derive constraints on the black hole mass and accretion
rate.  In \S5, we show results from detailed Monte Carlo simulations
of the thermal Comptonization and disk thermal reprocessing physics
that verify the empirical relations we used in \S3 and the SEDs we
computed in \S4.  In \S6, we discuss our results and their
implications.  We conclude in \S7.

\section{Observations of NGC 5548}
The simultaneous ground-based/\IUE/\Ginga\ observations of NGC 5548
cover a period Jan 1989 through July 1990.  The \IUE\ observations
themselves took place from 1988 Dec to 1990 July and are described in
Clavel et al.\ (1991) and Clavel et al.\ (1992).  These references
provide the continuum fluxes at the source rest frame wavelengths
$\lambda 1350$\AA, $\lambda 1840$\AA, and $\lambda 2670$\AA.  For the
fluxes at the latter two wavelengths, we make the corrections for
Fe{\sc ii} and Balmer continuum contamination as described by Clavel
et al.\ (1992; see also Wamsteker et al.\ 1990).  The optical fluxes
at $\lambda 5100$\AA\ were obtained via ground-based observations by
Peterson et al.\ (1991).  We correct for extinction for all these wave
bands and for starlight contamination in the optical band as described
in Magdziarz et al.\ (1998).

The \Ginga\ data are from the monitoring campaign reported by Nandra
et al.\ (1991) and were re-extracted by Magdziarz et al.\ (1998) to
obtain additional effective area at the higher energies ($\ga
15$~keV). The re-extracted data cover energies from 1.7 to 30 keV.  We
have fit the \Ginga\ data from the nine epochs that were simultaneous
with the \IUE\ observations using a power-law + reflection model for
the continuum components (the {\sc pexrav} model [Magdziarz \&
Zdziarski 1995] in {\sc xspec}), a Gaussian function for the
fluorescent iron K$\alpha$ emission line, and a Galactic absorbing
column fixed at $N_H = 1.65 \times 10^{20}$\,cm.  In addition, we
included an ionized absorber that is illuminated by a continuum with
the same spectral index as the cut-off power-law (the {\sc absori}
model [Done et al.\ 1992] in {\sc xspec}).  This latter component is
required to model the absorption at energies below $\sim 3$~keV that
cannot be accounted for by the Galactic absorbing column alone.  This
spectral model is similar to that used by Magdziarz et al.\ (1998) to
fit these same data. The difference is that we do not model the
so-called ``soft excess'' since we do not find significant residuals
below $\sim 2.5$~keV that would indicate the need for a separate
spectral component at these energies.\footnote{Magdziarz et al.\
(1998) were partly guided by non-simultaneous {\it ROSAT} data, which
we do not use here.  Only one epoch (7) of the \Ginga\ data shows
significant residuals above a power law at the softest X-ray
energies.}  Nonetheless, we find very similar X-ray spectral indices
and fluxes in the 2--10~keV band as do Magdziarz et al.\ (1998).  The
parameters from the fits to the \Ginga\ data along with the corrected
optical and UV fluxes are given in Table~\ref{IUE_Ginga_data}.

The simultaneous \EUVE, \ASCA, and \RXTE\ observations took place in
June--August 1998 and are described in detail in Chiang et al.\
(2000).  For uniformity's sake, we have used only spectral
parameters obtained from fitting \RXTE/\PCA\ data.  The \RXTE\
observations covered a greater period of time than did the \ASCA\
observations and overlapped more with the optical observations we
discuss below.  The \PCA\ response matrices used in the original
spectral analyses of these data were generated by the FTOOL utilities
{\sc pcarmf v3.5} and {\sc pcarsp v2.36} (Chiang et al.\ 2000).
However, the most recent versions of these programs as of September
2002, {\sc pcarmf v8.0} and {\sc pcarsp v8.0}, produce response
matrices that are sufficiently different from the earlier ones so that
the newly fitted spectral indices differ from those presented in
Chiang et al.\ (2000) by amounts that are $\sim 3$--4 times larger
than the 90\%~confidence level statistical errors.  In order to
illustrate the effect of the systematic uncertainties implied by these
changes, we have performed our analyses using the spectral parameters
obtained for both sets of response matrices.  In
Table~\ref{RXTE_data}, the 2--10~keV fluxes and spectral indices are
given for the ``1999 PCA responses'' and the ``2002 PCA responses''.
The former are taken directly from Table~6 of Chiang et al.\ 2000,
while the latter have been determined for this paper.  

In Table~\ref{RXTE_data}, the Compton reflection fractions, $R$, for
each epoch are also presented.  For the \RXTE\ data, there is clearly
a slight correlation between the reflection fraction and the photon
spectral index such that the harder power-law spectra for the new fits
are accompanied by weaker contributions from Compton reflection.
However, the changes in $R$ are not sufficiently large to account for
the changes in the underlying spectral index.

Simultaneous UV fluxes do not exist for the 1998 \EUVE/\ASCA/\RXTE\
observations, but we do have optical data at $\lambda 5100$\AA\ for
seven of the epochs which were kindly provided to us by M.~Dietrich
(Dietrich et al.\ 2001).  These data were extinction and starlight
corrected as described in Magdziarz et al.\ (1998).  The corrected
flux values are given in Table~\ref{RXTE_data}.  Unfortunately, the
optical flux measurements were not all strictly simultaneous with the
\RXTE\ observations.  For epoch 1.0, the flux listed in
Table~\ref{RXTE_data} was taken at JD$-2450000 = 980.751$, more than 8
hours after the \RXTE\ observation window.  For epochs 2.1--2.4, the
optical measurements were taken either at the very beginning or the
very end of the \RXTE\ integration intervals which lasted from $\sim
4$ to $\sim 17$ hours. Given the uncertainties in the optical fluxes
and the occasional sparseness of the coverage (see Dietrich et al.\
2001, their Fig.~4), interpretation of these data as being
simultaneous with the X-ray observations should be regarded with
caution.

\section{Semi-Analytic Calculations}

As in Papers~I and~II, we consider a fairly simple model for the
central regions of NGC~5548.  A razor thin, flat accretion disk
produces thermal emission that comprises the observed optical and UV
continuum; and a centrally located, spherical Comptonizing region
consisting of a plasma of hot electrons produces the power-law-like
X-ray continuum detected at energies $\sim 0.1$--$100$~keV.  The disk
is truncated at an inner radius $\rmin > 10\,GM/c^2$, the Comptonizing
electrons are assumed to have a uniform temperature $T_e$,
and the plasma sphere has a radius $r_s \sim \rmin$.  Thermal emission
from the disk enters the sphere and provides seed photons for the
Comptonization process.  In turn, X-rays from the sphere illuminate
the disk, heating it, and thus causing it to produce thermal emission
in excess of that resulting from local viscous dissipation.  The
feedback between the disk and sphere regulates the equilibrium
temperature of the Comptonizing electrons, and variations in the
geometry through changes in $\rmin$ and/or $r_s$ determine the
spectral properties of the X-ray, optical, and UV continua.  This sort
of geometry for thermal Comptonization in active galactic nuclei
(AGNs) and Galactic X-ray binaries has been explored in detail by a
number of authors (e.g., Shapiro, Lightman, \& Eardley 1976; Poutanen,
Krolik, \& Ryde 1997; Dove et al.\ 1997; Zdziarski, Lubi\'nski, \&
Smith 1999).

We model the thermal emission from the surface of the accretion disk
as
\begin{equation}
F_\disk = F_\visc + (1-a)F_\inc,
\label{Fdisk}
\end{equation}
where $a$ is the disk albedo.  The local flux produced by viscous
dissipation is
\begin{equation}
F_\visc = \frac{3}{8\pi} \frac{GM\Mdot}{r^3} 
            \left[1 - \pfrac{r_I}{r}^{1/2}\right],
\label{Fvisc}
\end{equation}
where $r_I$ is the radius of the innermost stable circular orbit.
Note that use of this expression assumes that the torque at the thin
disk truncation radius $\rmin$ is the same as that for a standard,
untruncated thin disk with zero torque at $r_I$.  We are also assuming that
the disk is stationary, except for the variable heating resulting from
the incident X-rays.  This incident X-ray flux
upon the disk is given as a function of radius by
\begin{equation}
F_\inc = \frac{3}{16\pi^2}\frac{L_x}{r_s^2}h\pfrac{r}{r_s},
\label{Finc}
\end{equation}
where $L_x$ is the X-ray luminosity of the Comptonizing plasma, and
the auxiliary function $h(r/r_s)$ is given by Zdziarski et al.\ (1999;
see also Paper~II).  This expression assumes that the X-ray emission
is uniform, isotropic, and optically thin throughout the plasma
sphere.  The disk emission as seen by an observer at a luminosity
distance $d_l$ and an inclination $i$ is given by
\begin{equation}
F_\nu = \frac{4\pi\cos i}{d_l^2}\frac{h\nu^3}{c^2}
        \int_\rmin^\infty\frac{r\,dr} {f_{\rm col}^4 [\exp(h\nu/k_B
        T_\disk\fcol) - 1]},
\label{Fnu}
\end{equation}
where the disk effective temperature is $T_\disk =
(F_\disk/\sigma_B)^{1/4}$, and $\fcol = \fcol(T_\disk)$ is a
temperature dependent color correction appropriate for AGN disks
(Paper~II; see also Hubeny et al.\ 2001).  In calculating $d_l$, we
assume a Hubble parameter of $H_0=67$~km~s$^{-1}$~Mpc$^{-1}$.

The function $h(r/r_s)$ can be approximated by
\begin{equation}
h\pfrac{r}{r_s} \simeq \frac{\pi}{4}\pfrac{r}{r_s}^{-3}
\label{h_approx}
\end{equation}
for $r \ga r_s$ (Zdziarski et al.\ 1999; see also Paper~I), and
therefore the disk thermal emission follows an $\sim r^{-3}$
distribution whether it results from viscous dissipation or from
reprocessing of the X-rays from the central plasma sphere, provided
that $r_s \sim \rmin \ga 10\,r_I$.  Under these circumstances, the
disk surface temperature can be modeled as $\tilde{T}_\disk \simeq T_0
(r/\rmin)^{-3/4}$.  Using this approximation, we have obtained
estimates of the disk inner truncation radius $\rmin$ and the
temperature $T_0$ at this radius by fitting the optical and UV data
with equation~(\ref{Fnu}).  Applying this procedure to the nine epochs
of the 1989--90 data, we have found the $\rmin$-values listed in
Table~\ref{Ginga_fits}.  These results depend sensitively on the disk
inclination, and that dependence affects the inferred values for the
other parameters describing the X-ray portion of the SEDs.  We will
explore the implications of this dependence in \S5.

The underlying relationship between the X-ray luminosity, the black
hole mass, and the accretion rate can be seen by using the
approximation for $h(r/r_s)$ (eq.~[\ref{h_approx}]) and the
aforementioned fits for $\rmin$ and $T_0$.  Since broad line region
reverberation mapping estimates of the central black hole mass in
NGC~5548 give $M_{\rm rm} = 6 \times 10^7$~\Msun\ (Peterson \& Wandel
2000), and $\rmin > 20\,GM_{\rm rm}/c^2$ for all nine of the 1989--90
epochs, we can neglect the term containing $r_I/r$ in the expression
for $F_\visc$ (eq.~[\ref{Fvisc}]).  Using these approximations, the
black hole mass and the accretion rate are related to the X-ray
luminosity $L_x$ and the radius of the plasma sphere $r_s$ by
\begin{equation}
GM\Mdot + (1-a)\frac{L_x r_s}{8} = \frac{8\pi\sigma_B T_0^4 r_\min^3}{3}.
\label{MMdot}
\end{equation}
Since we are assuming that the disk is stationary (eq.~[\ref{Fvisc}]),
the non-varying part of the optical/UV SED is modeled by the $\MMd$
term on the left-hand side of equation~(\ref{MMdot}), while any
optical/UV variability in the model must be caused by changes in the
thermal reprocessing component represented by the $L_x r_s$ term
and/or by changes in the truncation radius $\rmin$.

The values of $L_x$ and $r_s$ are also constrained by the observed
2--10 keV X-ray flux and spectral index, and by two energy balance
conditions.  The first energy balance condition follows from an
empirical relation between the X-ray spectral index and the Compton
amplification factor $A$, which is the ratio of the X-ray luminosity
and the luminosity of ``seed'' photons, $L_s$, entering the
Comptonizing plasma:
\begin{eqnarray}
\Gamma &=& \Gamma_0 (A - 1)^{-1/\delta}\\
       &=& \Gamma_0 \parens{\frac{L_x}{L_s} - 1}^{-1/\delta}.
\label{Gamma(A)}
\end{eqnarray}
The parameters $\Gamma_0 = 2.15$ and $\delta = 14$ have been found
from fits to Monte Carlo calculations of the thermal Comptonization
process (Malzac, Beloborodov, \& Poutanen 2001; see also \S4).  If
the seed photons consist only of thermal emission from the disk, then
$L_s$ is determined by the disk flux distribution $F_\disk(r)$ and by
the geometry of the disk and Comptonizing plasma through $r_s$ and
$\rmin$:
\begin{equation}
L_s = 4\pi\int_\rmin^\infty g(\rt; \rt_s) I(r) r\,dr,
\label{Ls}
\end{equation}
where $I(r) = F_\disk(r)/\pi$ is the disk surface brightness, $\rt
\equiv r/\rmin$, and $g(\rt; \rt_s)$ is defined in Paper~II and equals
the fraction of isotropic emission that enters the central plasma
sphere from a disk annulus $(\rt,\rt+d\rt)$.  Modeling the X-ray
spectrum as an exponentially cut-off power-law, $L_E \propto
E^{1-\Gamma} \exp(-E/E_c)$, where $E_c \equiv 2 \kT$, a value for
$L_x$ can be found using the 2--10~keV X-ray continuum flux,
$F_{210}$, and X-ray spectral index, $\Gamma$:
\begin{equation}
L_x = \xi 4\pi d_l^2 F_{210} \frac{Q(\Gamma,E_\min/E_c)}
      {Q(\Gamma,2\,{\rm keV}/E_c) - Q(\Gamma,10\,{\rm keV}/E_c)}.
\label{Lx}
\end{equation}
Here $E_\min = 0.01$~keV is an assumed nominal lower cut-off energy,
the precise value of which is not important for photon indices $\Gamma
< 2$, and
\begin{equation}
Q(\Gamma, x) \equiv \int_x^\infty t^{1-\Gamma} e^{-t} dt.
\end{equation}
Note that this expression is the incomplete gamma function.  The
factor $\xi$ in equation~(\ref{Lx}) parameterizes any anisotropy in
the thermal Compton emission.  If this emission were isotropic and the
plasma sphere were optically thin, then $\xi = 1$.  Because photons
from the first scattering order are preferentially scattered back
towards the disk, the X-ray intensity incident upon the disk can be
larger than the intensity seen by a distant observer (Paper~II).  Our
Monte Carlo calculations show that $\xi \simeq 1.5$ does a reasonable
job of modeling this anisotropy for observer inclinations $i \sim
30$--$45^\circ$ (see \S4).

The second energy balance condition is more global in nature and comes
from the expectation that the hot plasma is less efficient at
radiating accretion power than a standard optically thick accretion
disk would be in its place:
\begin{equation}
L_{\rm untr} \ga L_x + L_{\rm tr}.
\label{geb}
\end{equation}
The quantities $L_{\rm tr}$ and $L_{\rm untr}$ are the luminosities
resulting from viscous dissipation for a disk truncated at an inner
radius $\rmin$ and an ``untruncated'' disk extending to the innermost
stable circular orbit $r_I$, so that
\begin{equation}
L_{\rm tr, untr} \equiv 4\pi\int_{\rmin, r_I}^\infty r\,dr F_\visc(r).
\label{L_tr_untr}
\end{equation}
For a given value of $r_I$, equation~(\ref{geb}) is equivalent to
\begin{equation}
L_x < \eta\Mdot c^2 - L_{\rm tr},
\label{global_energy_balance}
\end{equation}
where the efficiency $\eta = 0.5$ for $r_I = GM/c^2$.\footnote{This is
obtained from equation~(\ref{L_tr_untr}).  We do not use the more
familiar value of $\eta = 0.42$ since we are neglecting general
relativistic effects throughout this work.}  We generally assume that
a firm upper limit to the X-ray luminosity is given by $\eta = 1$.  Of
course, this high a value would correspond to accretion efficiencies
that exceed both that of the standard thin disk and those expected for
other models such as advection dominated accretion flows (ADAFs;
Narayan \& Yi 1995).  However, even higher efficiencies are
conceivable if the spin energy of the central black hole can be tapped
(e.g., Krolik 1999; Gammie 1999; Wilms et al.\ 2001).

To summarize, we effectively have five relations constraining the
properties of the inner accretion flow.  We list them here in terms of
their functional dependence on the model parameters and the
observationally determined quantities:
\begin{mathletters}
\begin{eqnarray}
\MMd &=& \MMd(L_x, r_s, \rmin; F_{\rm OUV}), \\
\Gamma &=& \Gamma(L_x, L_s; \Gamma_0, \delta), \\
L_s &=& L_s(L_x, r_s, \rmin, \MMd), \\
L_x &=& L_x(T_e; F_{210}, \Gamma), \\
L_x & < & L_{\rm untr}(\MMd, r_I) - L_{\rm tr}(\MMd, \rmin).
\end{eqnarray}
\label{model_relations}
\end{mathletters}
The first relation represents equations~(\ref{Fdisk}--\ref{Fnu}) (or
approximately eq.~[\ref{MMdot}] by itself); and the remaining four
relations represent equations~(\ref{Gamma(A)}), (\ref{Ls}),
(\ref{Lx}), and~(\ref{global_energy_balance}), respectively.  In these
five relations, there are six unknowns\footnote{In fact, the disk
inclination $i$ is another unknown.  For clarity, we defer explicit
discussion of it until \S5 and assume $\cos i = 0.7$ for our analyses
of the NGC 5548 data unless otherwise indicated.}, $L_x$, $L_s$,
$T_e$, $r_s$, $\MMd$, and $r_I$; and we consider the remaining
quantities as being either directly measured from or constrained by
the observations: $F_{210}$ and $\Gamma$ from the X-ray data, and
$\rmin$ by the optical and UV data as represented by $F_{\rm OUV}$
(see eq.~[\ref{Fnu}]).  In order to obtain a unique solution to
equations~(\ref{model_relations}), additional information is
necessary.  Direct measurement of the roll-over energy $E_c$ would
provide the most useful constraints since that would provide estimates
of both $L_x$ and $T_e$, but instruments with the required sensitivity
in the relevant energy range are not yet available.  Reverberation
mapping measurements of the black hole mass can be used to estimate
$r_I$; but as we will see below, the standard value for NGC 5548,
$M_{\rm rm} = 6 \times 10^7$~\Msun\ (Peterson \& Wandel 2000), yields
accretion rates so low that equation~(\ref{global_energy_balance}) is
violated or that imply electron temperatures that are inconsistent
with the observed X-ray continua.

\section{Results}

\begin{figure}
\epsscale{0.8}
\plotone{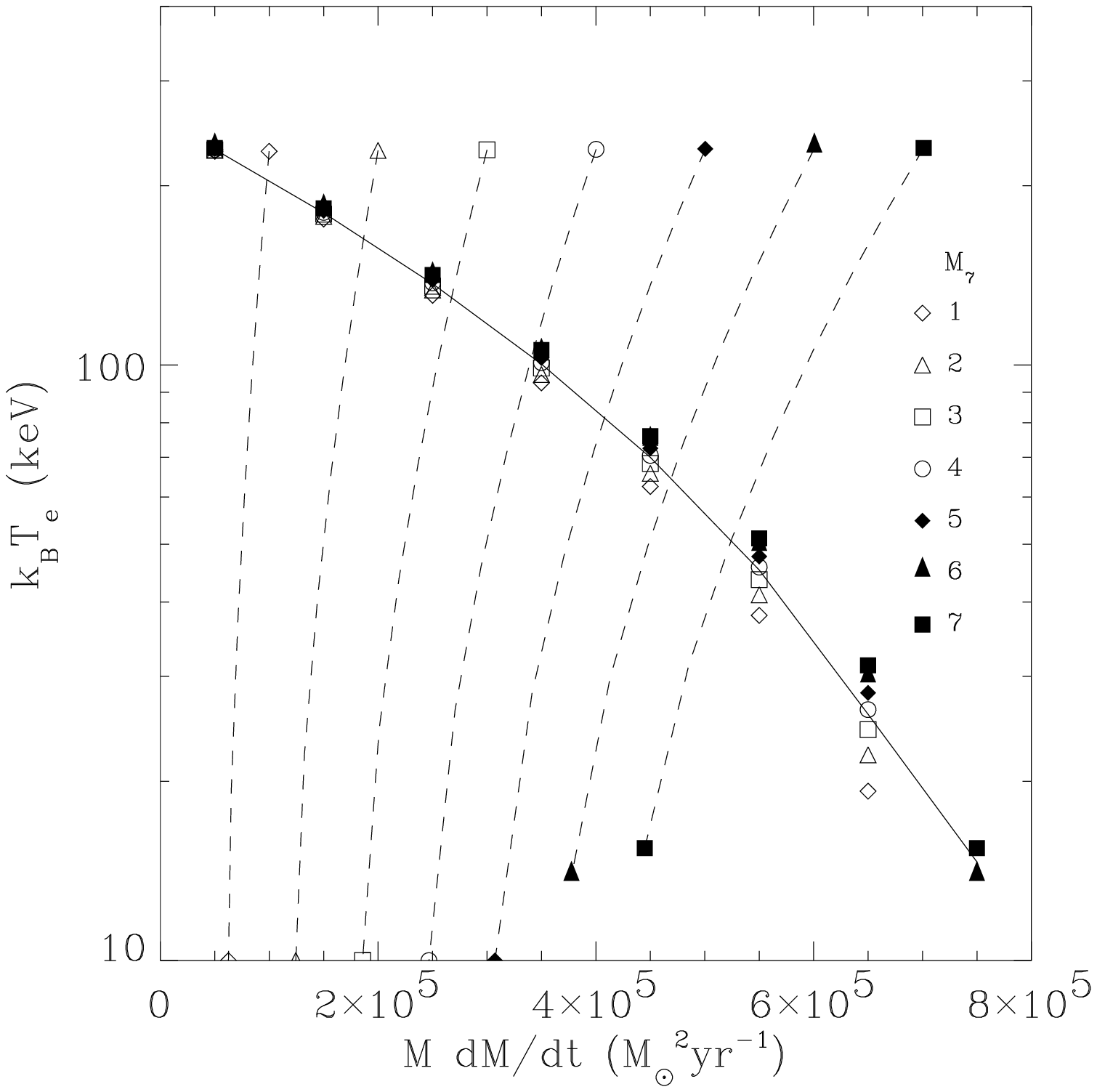}
\caption{Equilibrium electron temperature, $\kT$, versus $\MMd$
for the 1989--90 epoch~2 data, plotted as different clusters of points
from the upper left to the bottom right.  Each cluster corresponds to
a different chosen value of $\MMd = (0.5, 1.5, 2.5, 3.5, 4.5, 5.5,
6.5, 7.5)\times10^5$~\Msun$^2$~yr$^{-1}$.  Within each cluster the
different point symbols correspond to different black hole masses of
$M_7\equiv M/(10^7{\rm M}_\odot) = (1, 2, 3, 4, 5, 6, 7)$, as
indicated in the figure.  Note that $\kT$ depends primarily on $M\dot
M$, with only a weak dependence on $M$ arising because of the
resulting variation in $r_I$ (see eq. [\ref{Fvisc}]).  The solid curve
shows the relation between $\kT$ and $\MMd$ after we have averaged
over this weak mass dependence within each cluster of points.  The
dashed curves represent constraints from global energy balance, for
different values of black hole mass as indicated by the point symbols
at the ends of these curves.  Each of these dashed curves shows $\kT$
vs $\MMd_\min$, where $\dot{M}_\min \equiv (L_x+L_{\rm tr})/(\eta
c^2)$ and $\eta=0.5$.  For a given black hole mass, the intersection
of the dashed curve with the solid curve gives the minimum accretion
rate and maximum electron temperature (see \S4) that are consistent
with equations~\ref{model_relations}a--e.
\label{kT_vs_MMdot_example}}
\end{figure}

In order to get a sense for how the quantities $r_I$ and $\MMd$ can
affect the implied solutions, we have solved
equations~(\ref{model_relations}a--d) numerically over a grid of
values spanning $M = (1$--7)$\times 10^7$\Msun\ and $\MMd =
(0.5$--$7.5) \times 10^5$~\Msun$^2$~yr$^{-1}$, assuming $r_I =
GM/c^2$.  As an example, in Figure~\ref{kT_vs_MMdot_example}, the
values of $\kT$ obtained from these fits are plotted versus $\MMd$ for
the 1989--90 epoch~2 data.  Since $\rmin = 1.8 \times 10^{14}~{\rm cm}
> 10\,r_g$ over the entire range of masses, the $r_I/r$ term in
equation~(\ref{Fdisk}) plays a minor role in determining the disk flux
distribution, and equation~(\ref{MMdot}) roughly applies.
Consequently, the electron temperatures are nearly independent of $M$
and are determined mostly by the quantity $\MMd$.  This functional
dependence is shown in Figure~\ref{kT_vs_MMdot_example} by the various
symbols (corresponding to the different masses) and by the single
solid curve that connects the values of $\kT$ averaged over the masses
at each $\MMd$ value.  For each black hole mass, we compute the
minimum accretion rate that satisfies
equation~(\ref{global_energy_balance}),
\begin{equation}
\Mdot_\min \equiv \frac{L_x + L_{\rm tr}}{\eta c^2},
\end{equation}
where we have taken the accretion efficiency to be $\eta = 0.5$.  The
dashed curves show the fitted $\kT$ values plotted versus $\MMd_\min$
for the specified mass; solutions of
equation~\ref{model_relations}a--d that also obey
equation~\ref{model_relations}e will lie to the right of the dashed
curves.  Hence, the intersection of each dashed curve with the solid
curve gives the minimum value of $\MMd$ that satisfies the energy
balance condition (eq.~[\ref{global_energy_balance}]).  That
intersection point also corresponds to the maximum electron
temperature that can be achieved for the given mass.  For $M =
10^7$~\Msun, the electron temperature during epoch~2 must satisfy $\kT
\la 200$~keV; while for $M = 6 \times 10^7$~\Msun, it must satisfy
$\kT \la 60$~keV.

Several observations of the same source in different flux states can
place additional limits on the black hole mass $M$ and accretion rate
$\Mdot$, provided the latter is fixed as we are assuming by adopting a
stationary thin disk structure.  As we described above, we use the
optical and UV continuum fluxes to determine the disk inner radius,
$\rmin$, in each epoch.  In contrast to our previous studies of NGC
3516 and NGC 7469 (Paper~II), the optical and UV data for NGC 5548 do
seem to require variations in $\rmin$.  This is evidenced by the
relatively large optical/UV color changes occurring over these
observations (Table~\ref{IUE_Ginga_data}).  These observations are
separated in time by intervals ranging from two days to nearly one
year.  If variations in $\rmin$ result from viscous processes, then
any feasible observational time scale would be far too short to expect
significant changes.  However, if the $\rmin$-variations occur because
of thermal instabilities causing the disk to alternately ``puff up''
and then ``deflate'' near the innermost radius of the thin disk, then
such variations could occur on a thermal time scale:
\begin{equation}
t_{\rm th} \sim 10^6\,{\rm s} \pfrac{\alpha}{0.1}^{-1}
                \pfrac{\rmin}{2\times 10^{14}\,{\rm cm}}^{3/2}
                \pfrac{M}{10^7\,{\rm M}_\odot}^{-1/2},
\label{t_th}
\end{equation}
where $\alpha$ is the anomalous viscosity parameter.  This is roughly
consistent with the time scales of the $\rmin$-variations listed in
Table~\ref{Ginga_fits}, except perhaps for the 2-day interval between
epochs~4 and~5 that would require somewhat larger values of $\alpha$
and/or black hole mass than are indicated in equation~(\ref{t_th}).

\begin{figure}
\plotone{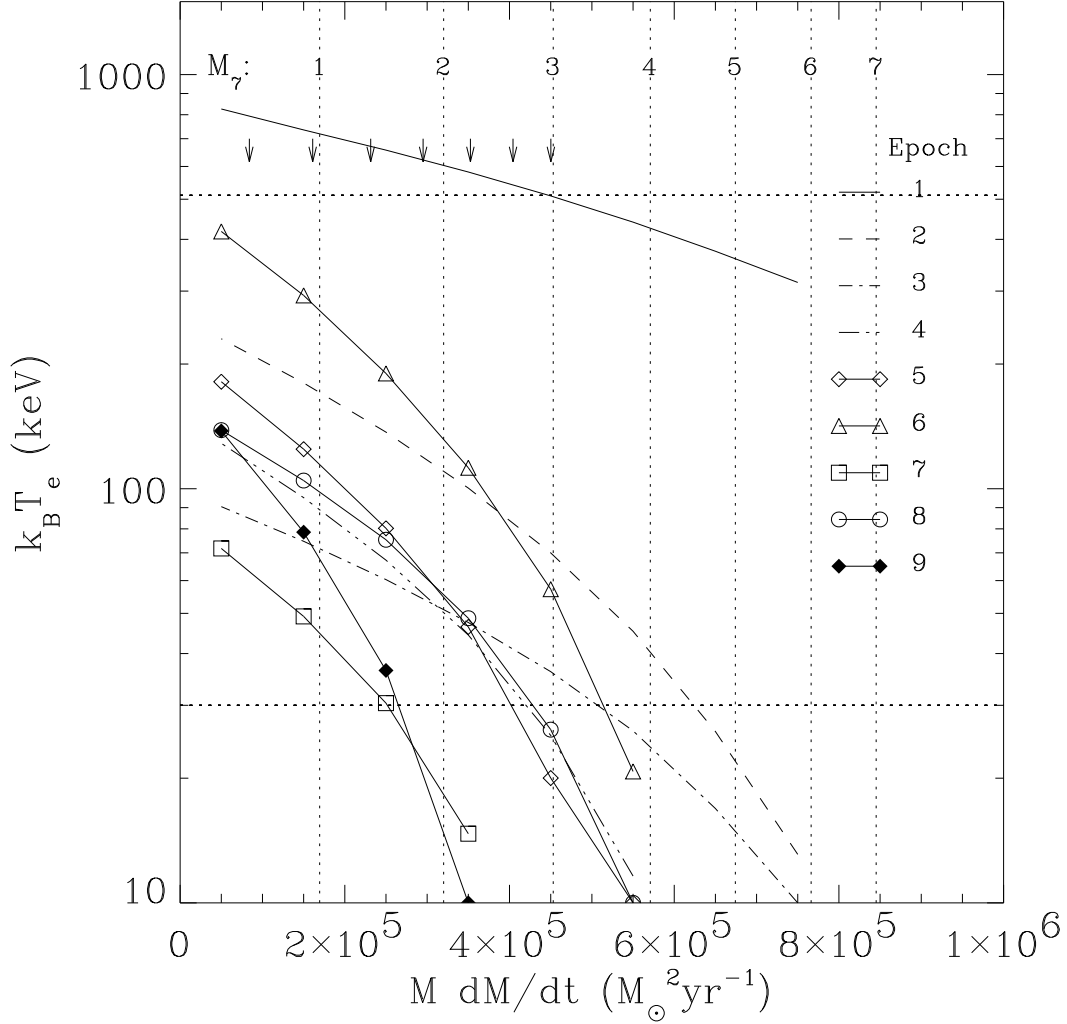}
\caption{Electron temperature versus $\MMd$ for all nine 1989--90
epochs.  The $\kT$ values for a given $\MMd$ have been averaged over
the seven black hole masses considered in
Figure~\ref{kT_vs_MMdot_example}.  The $\kT$-$\MMd$ curves for each
epoch are identified by the combination of linestyles and symbols as
shown in right-hand side of this figure.  The dotted vertical lines
indicate the minimum value of $\MMd$ for each mass for the epoch 1
data, which are the most constraining.  These lines correspond to
black hole masses $M = (1, 2, 3, 4, 5, 6, 7) \times 10^7$~\Msun,
ordered left-to-right and labeled near the top of the plot.  The
dotted horizontal lines show our assumed limits on the electron
equilibrium temperatures, $\kT_\min = 30$~keV and $\kT_\max = m_e c^2
= 511$~keV.  If we require $\kT \ga 30$~keV for all nine epochs, then
the $\kT$-$\MMd$ curves for epochs~7 and~9 constrain the black hole
mass to be $M < 2 \times 10^7$~\Msun\ and $\MMd \la 2.5 \times
10^5$~\Msun$^2$yr$^{-1}$.  The arrows near the top of the plot
indicate the minimum $\MMd$-values averaged over the constraints for
all nine epochs and correspond to $M = (1, 2, 3, 4, 5, 6, 7) \times
10^7$~\Msun, ordered left-to-right.  Therefore, taken as a whole, the
optical/\IUE/\Ginga\ data require $M \la 3\times 10^7$~\Msun.
\label{kT_vs_MMdot_Ginga}}
\end{figure}

In Figure~\ref{kT_vs_MMdot_Ginga}, we plot the $\kT$-$\MMd$ curves for
all nine of the 1989--90 epochs.  The vertical dotted lines indicate
the minimum $\MMd$ values for epoch 1 assuming masses $M = (1, 2, 3,
4, 5, 6, 7) \times 10^7$~\Msun, ordered left-to-right.  These lower
limits are greater than the corresponding ones found for the other
eight epochs, and so they represent the most restrictive minimum
values for $\MMd$.  Given these lower limits, a black hole mass of $M
= 6\times 10^7$~\Msun\ implies a minimum value of $(\MMd)_\min \simeq
7.5 \times 10^5$~M$_\odot^2$yr$^{-1}$ and requires rather low electron
temperatures for the eight other epochs, with $\kT \la 20$~keV in all
cases.  For temperatures that low, the thermal roll-over would be
detected in the X-ray spectra taken by instruments such as \Ginga\ and
\RXTE/\PCA, which have useful sensitivity up to $\sim 20$--$30$~keV
for the brighter type 1 Seyfert galaxies.  Of course, the degree to
which these observations could detect the effects of a low electron
temperature depends on the shape of the thermal Comptonization
continuum.  Although power-laws with exponential cut-offs are often
used to model these spectra, detailed thermal Comptonization
calculations, such as our Monte Carlo calculations (\S~5), generally
show sharper high energy cut-offs so that the thermal roll-over may
not be as evident at lower energies.  Furthermore, a strong Compton
reflection component can hide a low energy thermal roll-over much in
the same way it can mask a softer underlying power-law index.
Nonetheless, one can adopt a fairly conservative minimum value for
$T_e$ as an observational constraint.  For our purposes, we will
require $\kT > 30$~keV, which is indicated by the lower dotted
horizontal line in Figure~\ref{kT_vs_MMdot_Ginga}.  Since the
$\kT$-$\MMd$ curves decline fairly rapidly, a much lower minimum
temperature will not drastically alter the implied limits on $M$ and
$\Mdot$.  From Figure~\ref{kT_vs_MMdot_Ginga}, we see that the epoch~7
and~9 curves constrain $\MMd \la 2.5 \times 10^5$~M$_\odot^2$yr$^{-1}$
and the black hole mass to be $M < 2 \times 10^7$~\Msun.  Since we
expect the accretion efficiency to be less than maximal, we will take
$M = 1 \times 10^7$~\Msun\ and $\Mdot = 2 \times
10^{-2}$~\Msun~yr$^{-1}$ as nominal values hereafter.  Assuming $r_I =
GM/c^2$, this yields $L_{\rm untr} = \Mdot c^2/2 = 6 \times
10^{44}$~erg~s$^{-1}$ as our heuristic upper-limit for the bolometric
luminosity of this source (eq.~[\ref{global_energy_balance}]).  Note
that this is 60 percent of the Eddington luminosity at the chosen
black hole mass, which would be difficult to reconcile with an ADAF
interpretation for the inner accretion flow.

\begin{figure}
\plotone{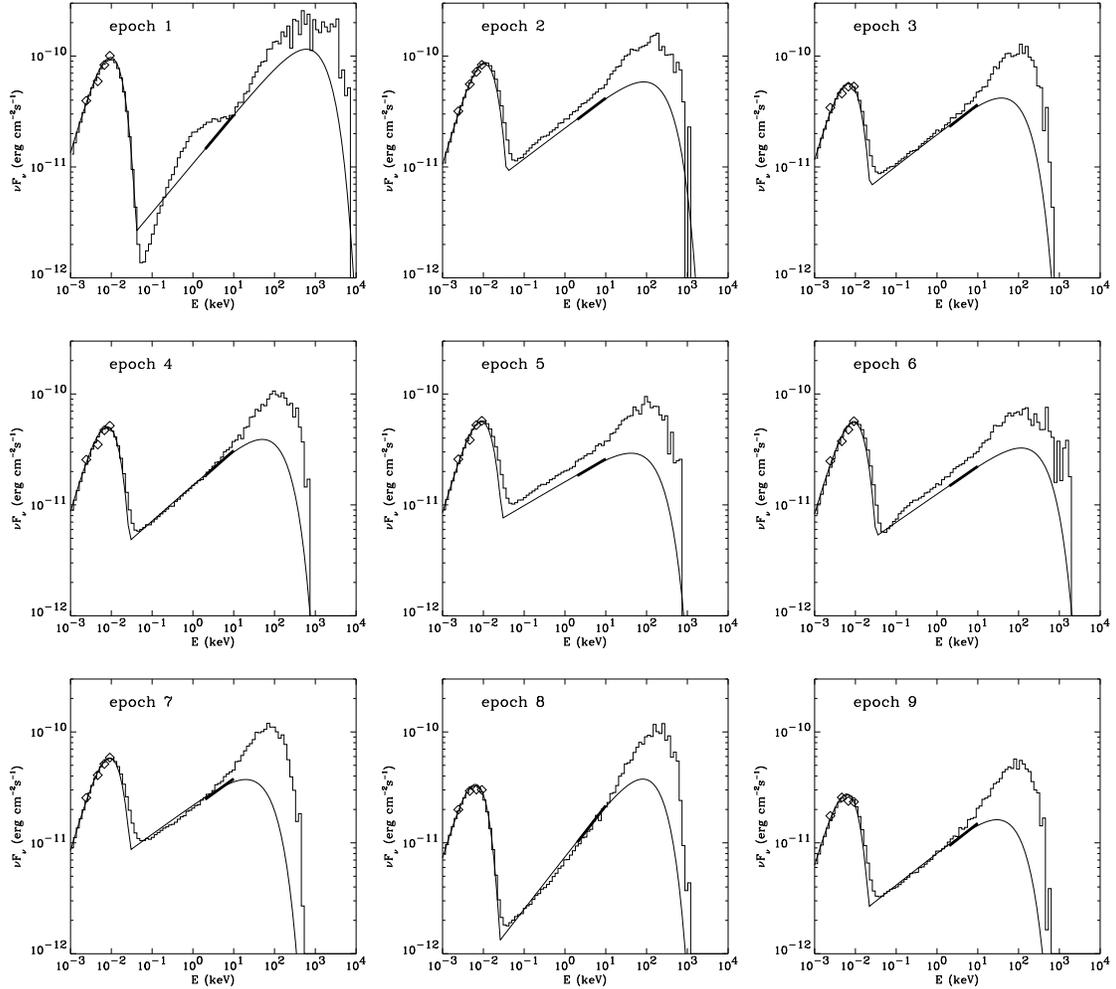}
\caption{Spectral energy distributions which have been ``fit'' to the
optical, UV, and X-ray data for the nine 1989--90 epochs.  The
optical/UV data are plotted as the open diamonds and the power-law
fits to the 2--10~keV \Ginga\ data are plotted as the thick line
segments.  The solid curves are the SEDs obtained from our
semi-analytic calculations described in \S3, and the histograms are
the thermal Comptonization/thermal reprocessing Monte Carlo results
using the parameters, $r_s$, $\rmin$, and $\tau_{\rm r} = \tau$, from
the semi-analytic calculations (see \S4).  All these SEDs have been
computed assuming $M = 1 \times 10^7$~\Msun\ and $\Mdot =
0.02$~\Msun~yr$^{-1}$.
\label{Ginga_SEDs}}
\end{figure}

The model parameters and SEDs which we find for the 1989--90 data
using these values are listed in Table~\ref{Ginga_fits} and shown in
Figure~\ref{Ginga_SEDs}, respectively.  In all cases we find that
$r_s/\rmin\sim1$, with this ratio being slightly greater or slightly
less than unity at different epochs.  Note that values less than unity
need not imply that there is a physical gap between the hot plasma and
the inner truncation radius of the disk.  Instead, it may indicate a
central accretion flow that is more flattened and aspherical compared
to the sphere we assume in the present model (see Paper~II).

The SEDs shown in Figure~\ref{Ginga_SEDs} often have more power in the
Comptonized X-ray component than in the optical/UV disk thermal
component.  This is still consistent with a radiatively inefficient
hot inner flow, because that efficiency should be compared to a
radiatively efficient disk at the same radius where most of the
accretion power is generated.  Indeed, the very fact that the X-rays
appear to take up such a large fraction of the bolometric power is
strongly suggestive of the accretion flow geometry we are considering
here.

\begin{figure}
\plotone{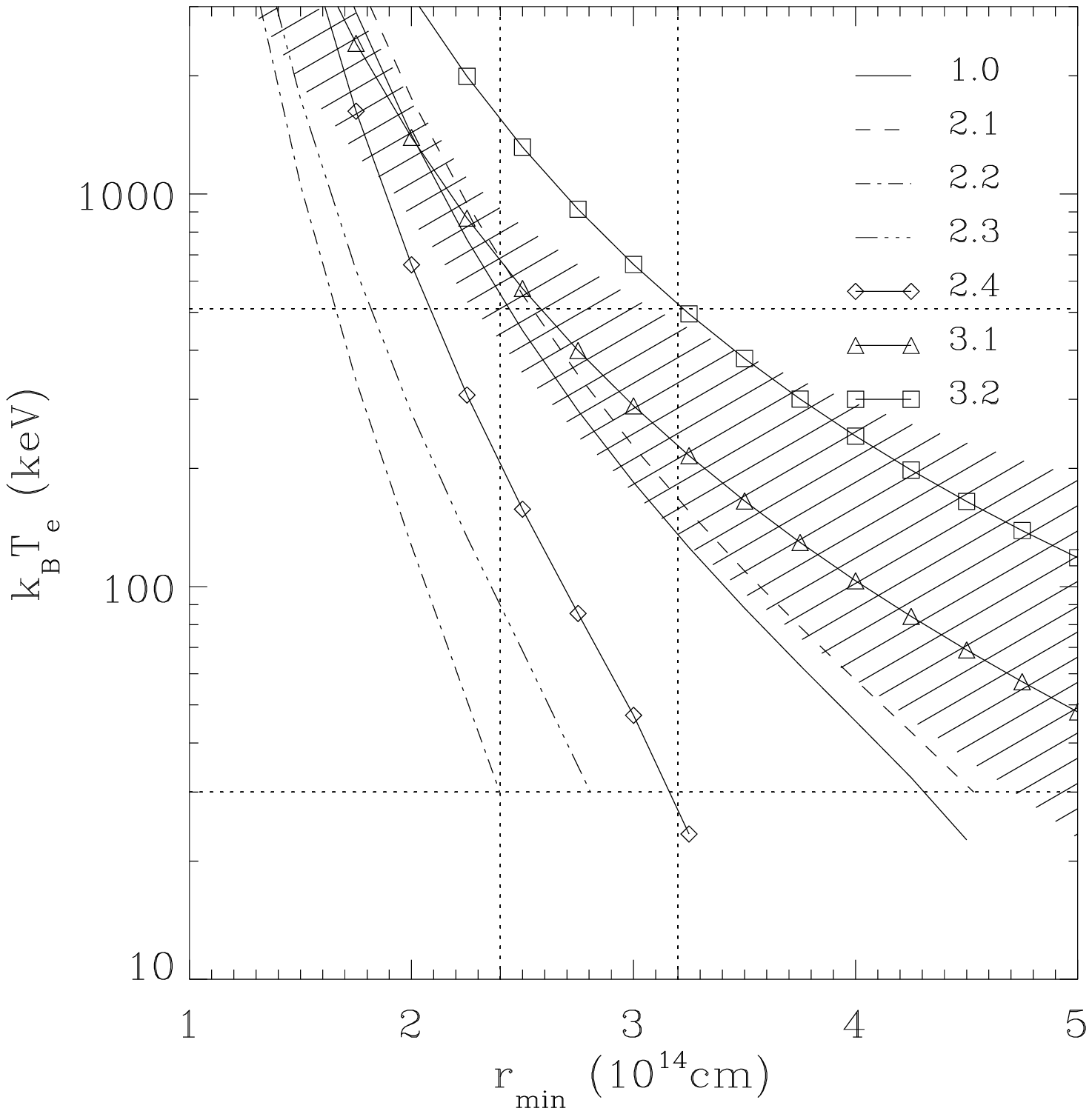}
\caption{$\kT$ versus $\rmin$ for the seven epochs of 1998
ground-based/RXTE data.  The mass and accretion rate have been fixed
at $M = 1\times10^7$~\Msun\ and $\Mdot = 0.02$~\Msun~yr$^{-1}$. The
horizontal dotted lines are our assumed upper and lower limits for the
electron temperature, $\kT_\max = 511$~keV and $\kT_\min = 30$~keV.
The individual curves have been obtained using the 1999 \RXTE/\PCA\
response matrices.  The lower limit $\kT_\min$ and the epoch~2.2 curve
imply a maximum disk inner radius of $\rmin \la 2.4 \times 10^{14}$~cm
(leftmost vertical dotted line).  By contrast, the upper limit
$\kT_\max$ and the epoch 3.2 curve require a {\em minimum} disk inner
radius, $\rmin \ga 3.2 \times 10^{14}$~cm (rightmost vertical dotted
line).  Taken together, these constraints imply that the disk inner
radius had to vary significantly over the course of the 1998
observations.  The hashed region shows the range of $\kT$ values at
each $\rmin$ that are obtained using the 2002 \RXTE/\PCA\ response
matrices.  For these fits, a single value of $\rmin$ in the range
(3.1--4.7)$\times 10^{14}$~cm may be taken for all seven epochs and
will be consistent with our nominal upper and lower electron
temperature limits.
\label{kT_vs_rmin_RXTE}}
\end{figure}

Applying our calculations to the 1998 observations is problematic
since we lack the UV data required to constrain the disk inner radius.
However, having ascertained $M$ and $\Mdot$ from the 1989--90 data, we
can explore the feasible range of disk inner radii for the 1998
epochs in light of our temperature constraint.  In
Figure~\ref{kT_vs_rmin_RXTE}, we plot the equilibrium electron
temperature versus $\rmin$ for the seven epochs of the 1998 data  for
both sets of response matrices.  For clarity, we have plotted the
individual curves that are obtained from analyses using the 1999 \PCA\
response matrices; while for the 2002 response matrices, we simply
show a hashed region that is bound by the minimum and maximum $\kT$
values at each value of $r_\min$.

Using the 1999 response matrices, the curve for epoch~2.2 requires
that $\rmin \le 2.4 \times 10^{14}$~cm ($= 160\,
G(M/10^7\,M_\odot)/c^2$) in order to satisfy $\kT > 30$~keV.  If we
assume that this upper limit for $\rmin$ applies to all of the 1998
observations, then the electron temperature for epoch~3.2 would be
$\kT \ga 1500$~keV (see Table~\ref{RXTE_fits}).  Since we do not
consider pair-production or the possibility of a non-thermal component
of relativistic electrons, another constraint we are tacitly imposing
in this model is that the electron temperatures should satisfy $\kT
\la m_e c^2$.  For temperatures much greater than this, the
thermalization time scales exceed the cooling time scales, and a
non-thermal electron component will arise (Coppi 1999).  The presence
of a significant non-thermal component would invalidate the energy
balance implied by the thermostatic interaction between the disk
reprocessed radiation and the Comptonization emission from the hot
plasma.  Therefore, in order to satisfy this implicit upper limit for
the temperature, the curves shown in Figure~\ref{kT_vs_rmin_RXTE}
suggest that the disk inner radii for these data span a range at least
as wide as $\rmin \simeq (2.4$--$3.2) \times 10^{14}$~cm.  However,
for the spectral parameters derived using the 2002 response matrices,
the disk inner radius can take {\em any single value} in the range
(3.1--$4.7) \times 10^{14}$~cm {\em for all seven epochs} and still be
consistent with electron temperatures that lie within our nominal
range of $30$--$511$~keV.  Since $r_s \sim \rmin$ for all these
calculations (see Tables~\ref{RXTE_fits} and~\ref{new_RXTE_fits}), it
is worth noting that, despite the changes in the \RXTE/\PCA\ response
matrices, all of these radii are roughly consistent with our estimates
of the size of the Comptonizing region that we derived from the
temporal lags between the soft (0.16~keV) and hard (2--10~keV) X-ray
light curves of the 1998 \EUVE/\ASCA/\RXTE\ observations (Chiang et
al.\ 2000).

\begin{figure}
\plotone{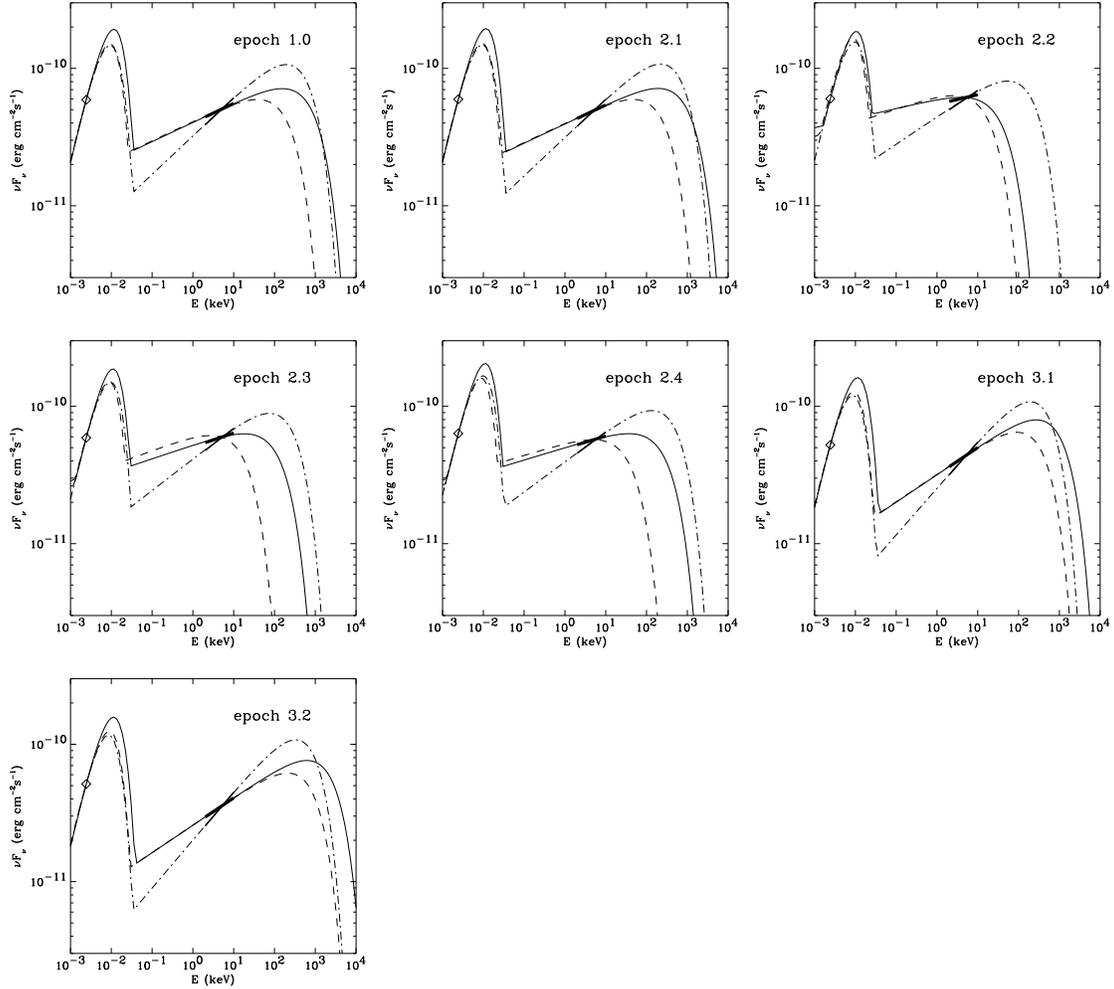}
\caption{Model spectral energy distributions for the 1998
ground-based/RXTE data.  Using the 1999 response matrices, SEDs
have been computed for two inner disk radii, $\rmin = 2.4\times
10^{14}$~cm (solid curves) and $\rmin = 3.2\times 10^{14}$~cm (dashed
curves).  The differences in thermal Comptonization cut-off energy and
the location and magnitude of the spectral peak in the UV band can be
clearly seen.  In epochs~2.2-2.4, the equilibrium electron
temperatures are all $< 30$~keV if one assumes a disk inner radius of
$\rmin = 3.2\times 10^{14}$~cm.  For an exponentially cut-off
power-law with $E_c = 2\kT$, these spectra differ substantially from
the single power-law component inferred for the 2--10~keV band from
the \RXTE/\PCA\ data (thick solid line segment). The SED
calculations using the 2002 response matrices are shown as dot-dashed
curves and a thinner line segment is plotted showing the 2--10~keV
power-law for these data.
\label{RXTE_SEDs}}
\end{figure}

In Figure~\ref{RXTE_SEDs}, we plot model spectral energy
distributions for the seven 1998 ground-based/\RXTE\ epochs.  Two of
the sets of curves have been computed using the parameters from the
1999 response matrix spectral fits, assuming $\rmin = 2.4 \times
10^{14}$~cm (solid curves) and $\rmin = 3.2 \times 10^{14}$~cm
(dashed).  The SEDs corresponding to the 2002 response matrix fits are
also shown (dot-dashed curves) and have been computed assuming $\rmin
= 3.2 \times 10^{14}$~cm.  The disk luminosities resulting from
viscous dissipation are $L_{\rm tr} = 1.0 \times 10^{43}$ and $0.7
\times 10^{43}$~erg~s$^{-1}$ for $\rmin = 2.4\times 10^{14}$~cm and
$3.2 \times 10^{14}$~cm, respectively.  Global energy balance,
equation~(\ref{global_energy_balance}), is obeyed in all cases.  The
SEDs shown in Figure~\ref{RXTE_SEDs} illustrate the dramatic effect
that the change in the response matrices have on the inferred broad
band X-ray continuum.  It is worth emphasizing, however, that the
2--10~keV continuum fluxes obtained using the two different sets of
response matrices are fairly consistent with one another (see
Table~\ref{RXTE_data}).  Consequently, the largest differences in the
model SEDs occur at energies that lie well outside the range directly
measured by \RXTE/\PCA.

\section{Monte Carlo Calculations}

We have implemented a three dimensional Monte Carlo calculation of the
thermal Comptonization/thermal reprocessing model we have described in
the previous section.  A more detailed description appears in the
Appendix and in Chiang \& Blaes (2002).  In this section, we compare
results from our Monte Carlo calculations with the empirical relations
we have used in the previous section and with the spectral energy
distributions obtained using those semi-analytic methods.

\begin{figure}
\plotone{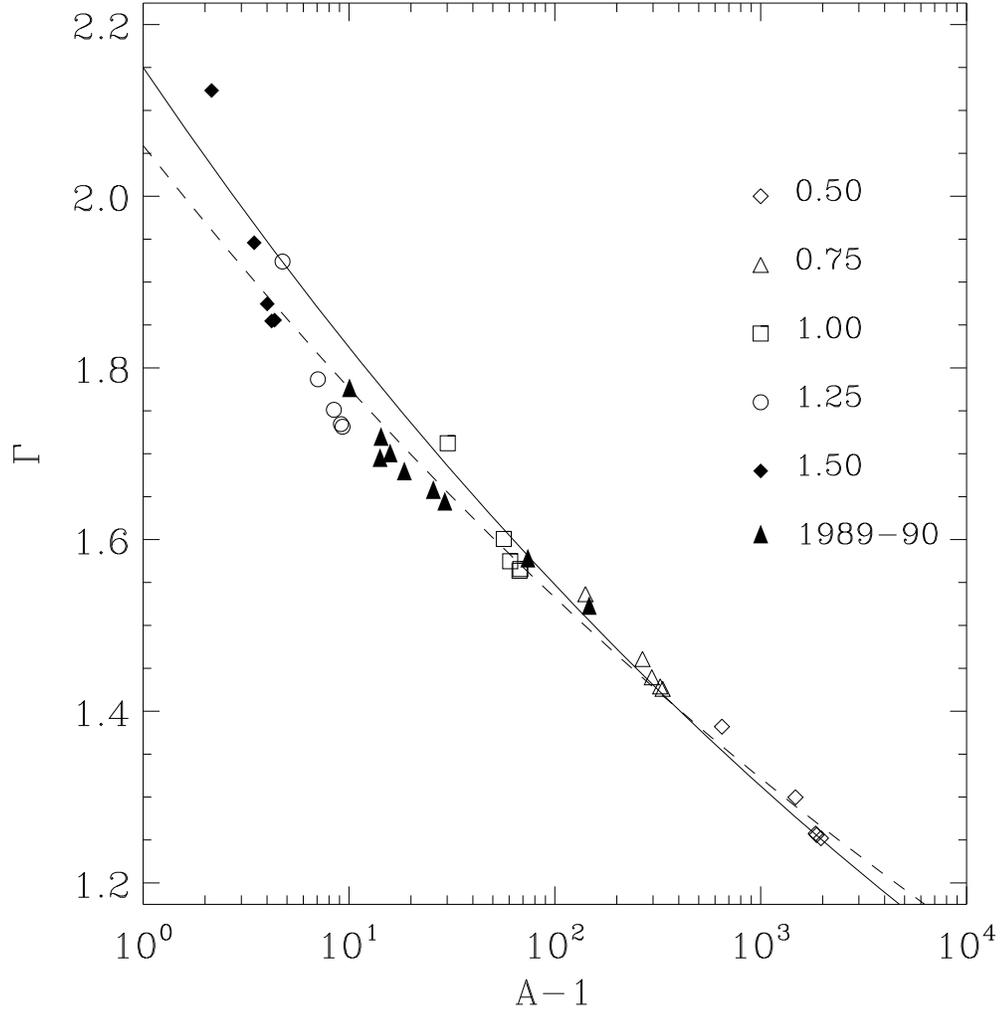}
\caption{Photon spectral index versus Compton amplification factor
minus unity.  The filled triangles are from the Monte Carlo
calculations shown in Figure~\ref{Ginga_SEDs} that were computed using
the parameters for the 1989--90 data. 
The other symbols correspond to separate sets of Monte Carlo calculations for
geometries with values of $r_s/r_{\rm min}=0.5$, 0.75, 1, 1.25, and 1.5, as
indicated in the figure.  Each set of Monte Carlo calculations had
$M=2\times10^7$~M$_\odot$ and different values of $\dot M=0$, 0.01, 0.02, 0.04,
and 0.08 M$_\odot$~yr$^{-1}$, with larger $\dot M$ values corresponding to
larger values of spectral index $\Gamma$.  (For larger values of $\dot M$,
smaller fractions of soft photons from thermal reprocessing are required,
and so softer X-ray spectra are produced.)
The solid
curve is the parameterization found by Malzac et al.\ (2001) which has
$\Gamma_0 = 2.15$ and $\delta = 14$ (see eq.~[\ref{Gamma(A)}]),
while the dashed curve is the best fit to the Monte Carlo data
yielding $\Gamma_0 = 2.06$ and $\delta = 15.6$.
\label{Gamma(A)_plot}}
\end{figure}

In Figure~\ref{Gamma(A)_plot}, we plot the X-ray continuum spectral
index versus the amplification factor, $A = L_x/L_s$.  The individual
data points are from our Monte Carlo calculations, the solid curve is
the parameterization of Malzac et al.\ (2001; see
eq.~[\ref{Gamma(A)}]), and the dashed curve is a best fit to the Monte
Carlo data with $\Gamma_0 = 2.06$ and $\delta = 15.6$.  The filled
triangles represent the Monte Carlo calculations that produced the
histogrammed SEDs shown in Figure~\ref{Ginga_SEDs}; we discuss these
simulations in greater detail below.  The other symbols represent
simulations that were computed for a black hole mass of $M = 2\times
10^7$\Msun, a disk inner radius $\rmin = 2\times 10^{14}$~cm, an
observer inclination of $i = 30^\circ$, five mass accretion rates
$\Mdot = 0, 0.01, 0.02, 0.04$, and $0.08$~\Msun~yr$^{-1}$, and five
sphere radii $r_s = 0.5, 0.75, 1.0, 1.25,$ and $1.5 \times 10^{14}$~cm
for a total of 25 spectra.

\begin{figure}
\plotone{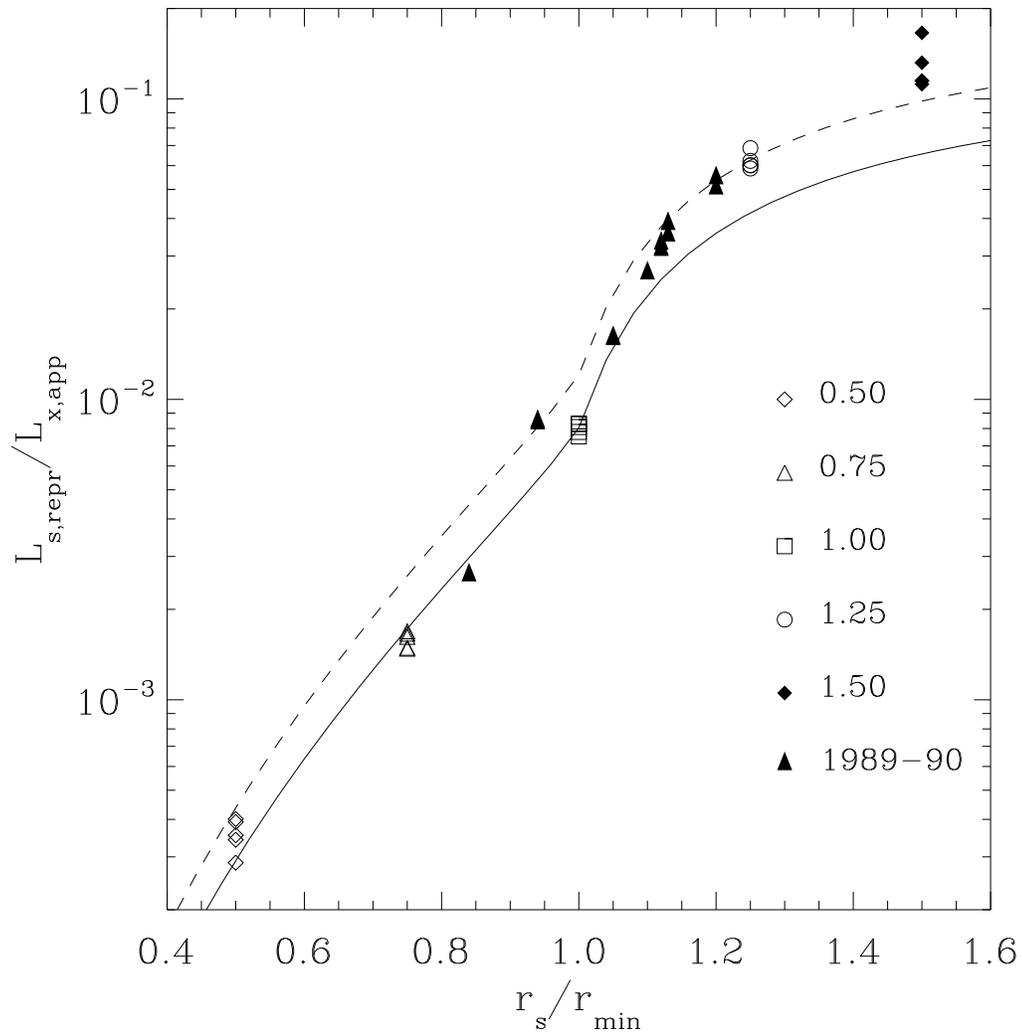}
\caption{The ratio of the luminosity of thermally reprocessed seed
photons, $L_{s, {\rm repr}}$, to the apparent X-ray luminosity,
$L_{x,{\rm app}}$, versus the ratio of Comptonizing sphere radius,
$r_s$, to disk inner radius, $\rmin$.  The data points are the Monte
Carlo calculations, and the various symbols correspond to the
same datasets, $r_s/\rmin$ and \Ginga\ data, as in
Figure~\ref{Gamma(A)_plot}.  The solid curve is the relation obtained
assuming the thermal Compton emission is isotropic.  The dashed curve
is the solid curve multiplied by $\xi = 1.5$.
\label{Ls_vs_rs}}
\end{figure}

In Figure~\ref{Ls_vs_rs}, we plot the ratio $L_{s,{\rm
repr}}/L_{x,{\rm app}}$ versus $r_s/\rmin$ for these same two sets of
simulations.  Here $L_{s,{\rm repr}}$ is the seed photon luminosity
produced by thermal reprocessing (i.e., the total $L_s$ minus the
portion produced by viscous dissipation), and $L_{x,{\rm app}}$ is the
{\em apparent} thermal Comptonization luminosity, i.e., the luminosity
one would compute from equation~(\ref{Lx}) for $\xi = 1$.  The solid
curve is the relation one would find if the thermal Comptonization
luminosity were indeed isotropic so that the flux incident upon the
disk is given by equation~(\ref{Finc}).  The dashed curve is the solid
curve multiplied by $\xi = 1.5$, the value we adopted to model the
anisotropy of the thermal Comptonization emission.  This factor also
takes into account differences in the albedo of the disk.  For the
semi-analytic calculations, we assumed a constant albedo of $a =
0.15$, while the disk albedos found in the Monte Carlo calculations
vary with the incident spectrum, ranging from $a \simeq 0.2$ for
$\Gamma \simeq 1.9$ to $a \simeq 0.4$ for $\Gamma \simeq 1.5$.  As can
be seen from Figure~\ref{Ls_vs_rs}, for $r_s/\rmin \ga 1.1$, a value
of $\xi = 1.5$ is appropriate, while for lower values, which
correspond to weaker coupling between the hot plasma and cold disk and
therefore produce harder spectra, a value of $\xi \simeq 1$ should
probably be used.

In Figure~\ref{Ginga_SEDs}, the histograms are the Monte Carlo SEDs we
computed using the values of $M$, $\Mdot$, $r_s$, and $\rmin$ found in
\S3 for the 1989--90 data.  The equilibrium electron temperatures we
obtain from our Monte Carlo calculations are systematically larger by
$\sim 20$\% than those from our semi-analytic methods.  Now, in the
Monte Carlo simulations, we have set the radial Thomson depth,
$\tau_{\rm r}$, of the Comptonizing sphere to be equal to the Thomson
depth, $\tau$, found from the observed spectral index, the electron
temperature from the semi-analytic method, and the empirical relation
(Beloborodov 1999),
\begin{equation}
\Gamma = \frac{9}{4}y^{-2/9},
\end{equation}
where the Compton $y$-parameter is 
\begin{equation}
y \equiv \tau(1 + \tau)4\theta_e(1+4\theta_e),
\end{equation}
and $\theta_e \equiv \kT/m_e c^2$.  Since the spectral index is set by
the geometric factors, $r_s$ and $\rmin$, $\Gamma$ is independent of
the Thomson depth of the plasma.  However, the amplitude of the
thermal Comptonization continuum is determined by the amount of seed
photons {\em scattered} in the Comptonizing region and so should be
roughly proportional to $1 - e^{-\tau_{\rm r}}$.  Therefore, the
mis-match between the continuum levels of the semi-analytic and Monte
Carlo calculations, most clearly seen in the epoch~5 results, can be
reduced by adjusting the radial Thomson depth of the plasma sphere.
This will, of course, result in different electron temperatures as
well.

For the epoch~1 data, the shape of the Monte Carlo SED is more
difficult to reconcile with the exponentially cut-off power-law used
in our semi-analytic method because for sufficiently high electron
temperatures the individual scattering orders are partially resolved
in the X-ray spectra.  This illustrates another, observationally-based
motivation for an upper limit on the electron temperature.  If the
mean separation in energy between the peaks of individual scattering
orders exceeds their width, these orders will be partially resolved in
the resulting thermal Comptonization continuum and will appear as a
number of spectral breaks and/or local maxima and minima (e.g., Skibo
\& Dermer 1995).  Empirically, we find from our Monte Carlo
calculations that an upper limit of $\kT \la m_e c^2$ is required to
prevent these features from appearing.

\section{Discussion}

The wide range of electron temperatures, $\kT \sim 40$--$700$~keV,
that we found for the ground-based/\IUE/\Ginga\ data is driven by the
relatively broad range of optical/UV fluxes (see
Table~\ref{IUE_Ginga_data}).  In Paper~II, we obtained similar results
for the NGC~7469 monitoring observations, which also had large
optical/UV variations.  In that work, we were able to narrow the range
of temperature changes that was required by allowing the X-ray
spectral indices to vary within their 1-$\sigma$ uncertainties.  For
the NGC~5548 ground-based/\IUE/\Ginga\ data, the most conspicuous out-lier
is the fit to the epoch~1 data.  Those data yield an electron
temperature that is at least three times larger than the temperatures
found for the other eight epochs.  The epoch~1 data also have an
unusually hard X-ray spectral index, $\Gamma = 1.56$.  This value lies
at the very low end of the range of spectral indices that are typical
for type 1 Seyfert galaxies (Zdziarski, Poutanen, \& Johnson 2000).
Furthermore, it is substantially smaller than the indices of the other
\Ginga\ epochs (with the notable exception of epoch~8).  However, as we
noted in Paper~II, softer spectra have less X-ray flux at energies $>
10$~keV than do harder spectra for a given value of $\kT$ and a fixed
2--10 keV flux.  In order to reduce $\kT$ for epoch~1, we would either
have to make the X-ray spectrum even harder, i.e., use $\Gamma <
1.56$, or take a larger value of the 2--10~keV flux.  Setting $F_{210}
= 3.73 \times 10^{-11}$~erg~cm$^{-2}$s$^{-1}$ for epoch~1, which
corresponds to a 10\% increase in 2--10~keV flux over the fitted value
(see Table~\ref{IUE_Ginga_data}), we obtain a uniform 20\% reduction
in the equilibrium electron temperature for all values of $\MMd$
considered.  The resulting changes in the inferred minimum accretion
rates are even smaller: for $M = 7\times 10^7$~\Msun, $\dot{M}_\min$
is reduced by 13\%, while for $M = 2\times 10^7$~\Msun, the reduction
is less than 4\%.

\begin{figure}
\epsscale{0.5}
\plotone{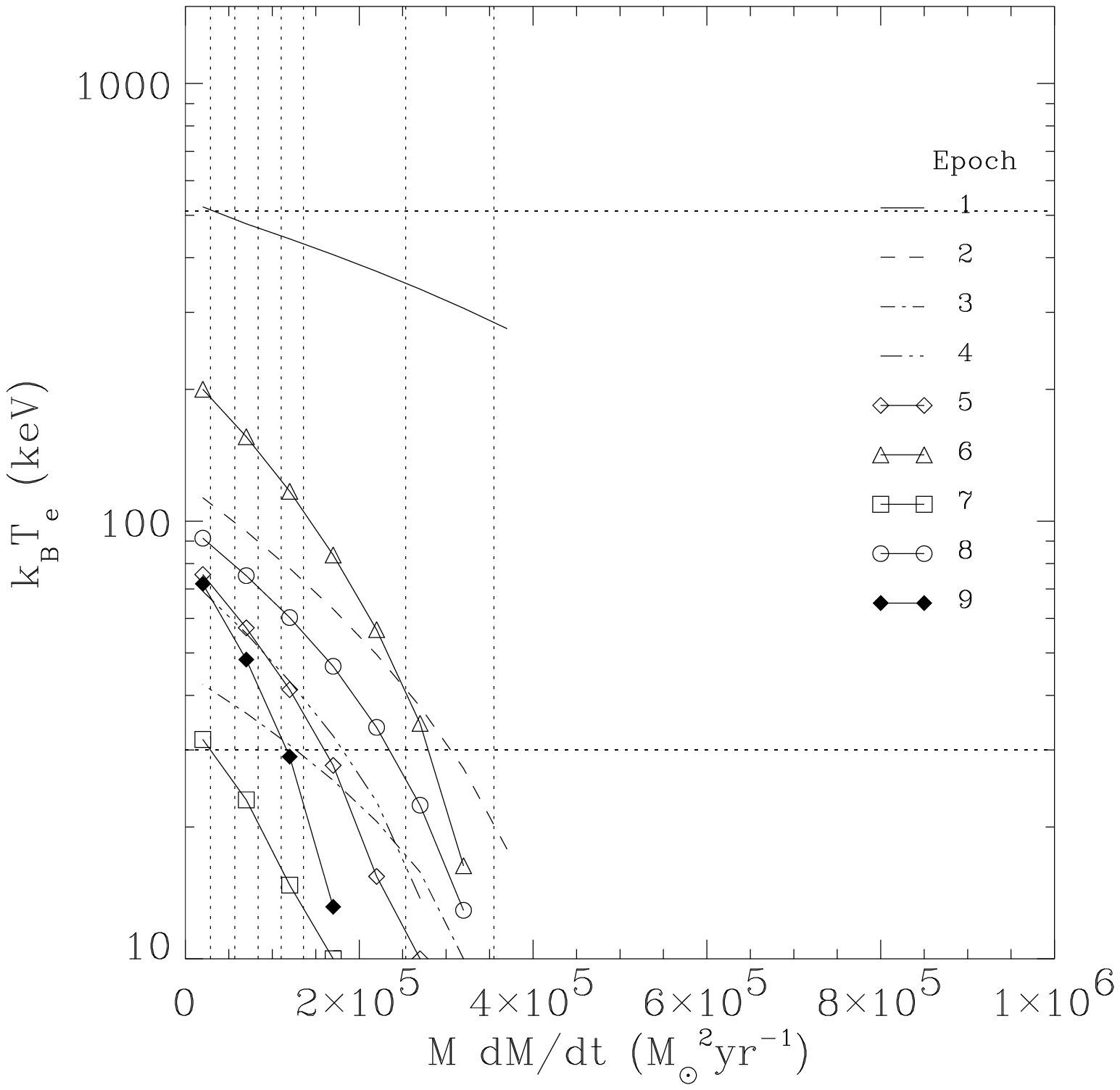}
\plotone{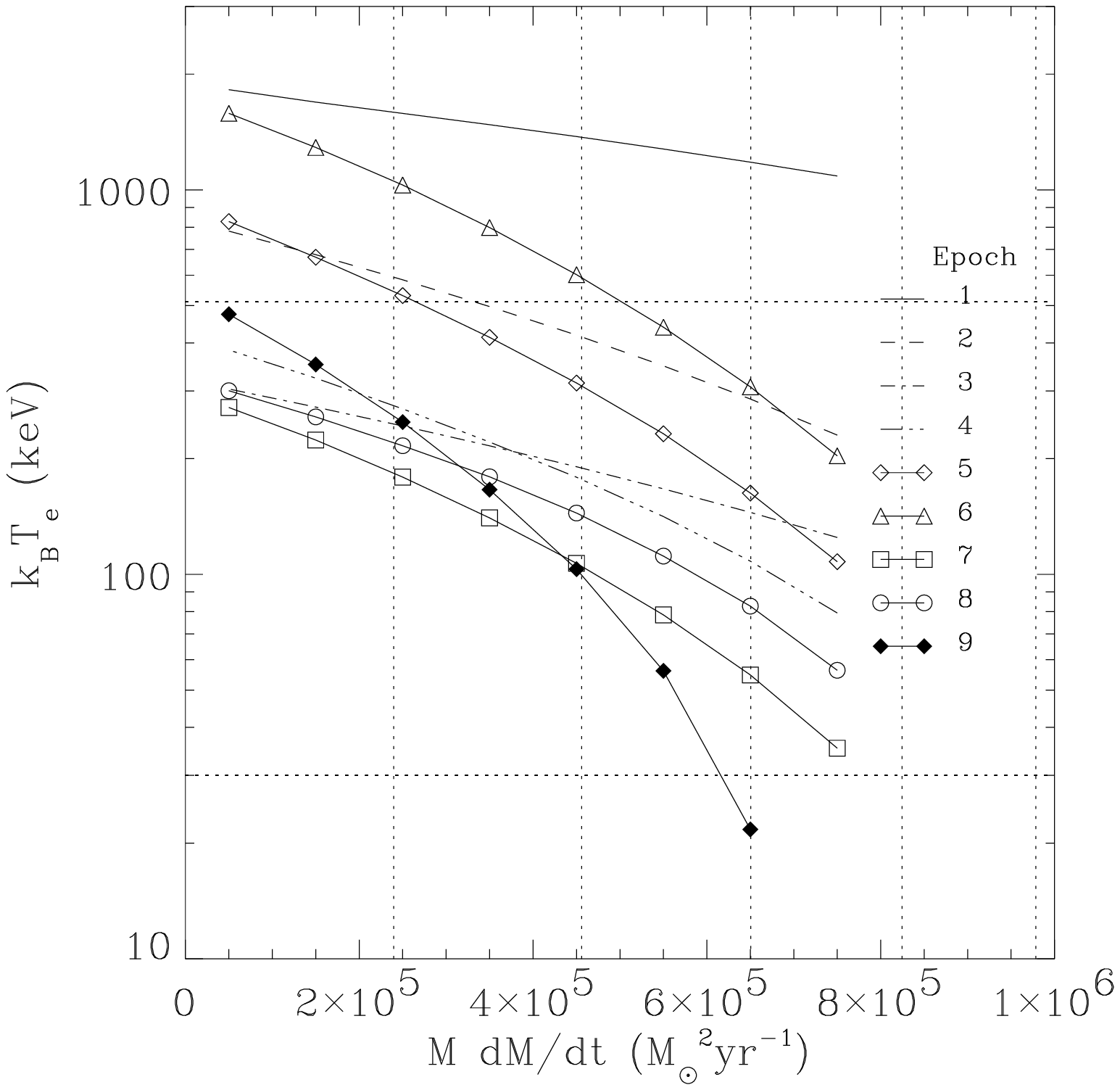}
\caption{$\kT$ versus $\MMd$ for two different observer inclinations,
$\cos i = 0.866$ (upper) and $\cos i = 0.5$ (lower).  These plots
illustrate the sensitivity of the limits on the black hole mass and
accretion rates to the assumed disk inclination.  For $\cos i =
0.866$, the dotted vertical lines are the minimum $\MMd$-values for
black hole masses $M = 0.2, 0.4, 0.6, 0.8, 1, 2, 3 \times 10^7$~\Msun,
ordered left-to-right, while for $\cos i = 0.5$, the dotted vertical
lines correspond to $\MMd$ limits for $M = 1, 2, 3, 4, 5 \times
10^7$~\Msun.  In both cases, these limiting values are from the
epoch~1 curve.  The horizontal dotted lines indicate
$\kT_\min=30$~keV and $\kT_\max = 511$~keV.
\label{kT_vs_MMdot_30_60}}
\end{figure}

For the preceding analyses of the NGC 5548 data we have assumed an
observer inclination of $i = 45^\circ\!\!.5$ ($\cos i = 0.7$).
Our calculations are surprisingly sensitive to this value.  Since the
optical/UV spectral shape for a given epoch is determined largely by
the temperature of the disk at the innermost radius, $T_0$ is
independent of $\cos i$.  The disk flux is proportional to $\sim \cos
i\,r_\min^2 T_0^4$, so that for a given set of optical/UV data, we
have $\rmin \propto (\cos i)^{-1/2}$.  If the disk emission is
dominated by the thermally reprocessed component, then the geometry is
determined by the X-ray spectral index (Zdziarski et al.\ 1999), and
we have $r_s \propto \rmin \propto (\cos i)^{-1/2}$, $L_s \propto
T_0^4 r_\min^2 \propto (\cos i)^{-1}$, and $L_x \propto (\cos
i)^{-1}$.  Inserting these relations into equation~(\ref{MMdot}), we
obtain
\begin{equation}
G\MMd = \frac{8\pi\sigma_B T_0^4 r_{{\rm min},0}^3}{3(\cos i)^{3/2}}
        - \frac{L_{x,0} r_{s,0}(1-a)}{8(\cos i)^{3/2}},
\label{MMdot_cos_i}
\end{equation}
where $L_{x,0}$, $r_{\min,0}$, and $r_{s,0}$ are fiducial values
evaluated for $\cos i = 1$.  The implied values of $\MMd$ will
therefore be larger for smaller $\cos i$.  Similarly, the global
energy balance condition (eq.~[\ref{global_energy_balance}]) allows us
to obtain a ($\cos i$)-scaling for $\Mdot$:
\begin{equation}
\eta \Mdot c^2 > \frac{4\pi  \sigma_B T_0^4 r_{\min,0}^2}{\cos i} +
                 [1 - (1-a)R]\frac{L_{x,0}}{\cos i},
\end{equation}
where $R$ is the fraction of the X-ray luminosity intercepted by the
disk.  Taking the minimum value for $\Mdot$ in order to have a maximum
value for the black hole mass, this expression and
equation~(\ref{MMdot_cos_i}) imply $M \propto (\cos i)^{-1/2}$ and
$\Mdot \propto (\cos i)^{-1}$.  Now, the X-ray observations and our ad
hoc limits for the electron temperature fix the range of thermal
Comptonization roll-over energies, and therefore they essentially fix
the $L_x$ values for the 1989--90 data as a whole (see
Figs.~\ref{kT_vs_MMdot_Ginga} and~\ref{kT_vs_MMdot_30_60}), so the
second term on the right-hand side of equation~(\ref{MMdot_cos_i})
will, in practice, go as $(\cos i)^{-1/2}$.  Nevertheless, the general
trend for $M$ and $\Mdot$ to be larger for smaller values of $\cos i$
still holds.  In fact, we find for these data that each term on the
right-hand side of equation~(\ref{MMdot_cos_i}) is much larger than
the left-hand side, so given the different effective scalings with
$\cos i$ of those two terms, the value of $\MMd$ is quite sensitive to
the assumed inclination.

Figure~\ref{kT_vs_MMdot_30_60} illustrates more concretely how much
the feasible ranges of $M$ and $\Mdot$ depend on the choice of the
observer inclination; we show plots of $\kT$ versus $\MMd$ for $i =
30^\circ$ and $i = 60^\circ$ for the 1989--90 data.  If we take our
minimum electron temperature limit seriously, then for $i = 30^\circ$,
we find $M \la 2 \times 10^6$~\Msun\ and $\MMd \la 2 \times
10^4$~M$_\odot^2$yr$^{-1}$.  By contrast, for $i = 60^\circ$, we find
$M \la 3 \times 10^{7}$\Msun\ and $\MMd \la 6 \times
10^5$~M$_\odot^2$yr$^{-1}$. In the latter case, an electron
temperature of $\kT \simeq 1200$~keV is implied for epoch~1, which
well exceeds our upper temperature limit.  For the former case,
although the electron temperatures fall within our adopted limits of
30~keV $< \kT < m_e c^2$, the implied black hole mass is a factor of
30 smaller than the reverberation mapping estimate.  Therefore, as a
compromise, we have taken $i \simeq 45^\circ$ as our nominal value
which does yield a somewhat high electron temperature ($\kT \simeq
700$~keV) for epoch~1, but also gives a mass upper limit ($M < 2
\times 10^7$~\Msun) that is not in too great conflict with the
reverberation mapping estimates (see Krolik 2001).
Note that extraction of black hole spin energy, which we have not
considered here, would ease the upper limits on the black hole mass
by allowing, in principle, accretion efficiencies in excess of unity.

Even assuming a mass as large as the reverberation mapping estimates,
our fits to the optical/UV continua give $\rmin > 20\,GM/c^2$.  This
is marginally inconsistent with the range of disk inner radii of
$\rmin = 7.5$--$15\,GM/c^2$ found by fitting a relativistically
broadened iron line to the 1998 \ASCA\ data (Chiang et al.\ 2000; see
also Mushotzky et al.\ 1995).  If $M \la 2 \times 10^7$~\Msun, then
our results would be in serious disagreement with the detection of a
relativistically broadened line.  This conclusion relies on the
1989--90 epoch~1 data providing the relevant limits on $\MMd$.  It is
clear from Figures~\ref{kT_vs_MMdot_Ginga} and~\ref{Ginga_SEDs} that
the SED during this epoch was unusual, and the source may have been in
a transient state for which the X-ray luminosity briefly exceeded the
limit imposed by equation~(\ref{global_energy_balance}).  Averaging
over the nine 1989--90 epochs, we obtain a much lower set of
$\MMd$-limits than we find from epoch~1 alone.  These are indicated in
Figure~\ref{kT_vs_MMdot_Ginga} by the arrows along the top of the
plot.  We still require that $\MMd \la 2.5 \times
10^5$~M$_\odot^2$yr$^{-1}$ in order to have reasonable electron
temperatures for epochs~7 and~9; and so these averaged limits imply $M
\la 3 \times 10^7$~\Msun.  If we take $M = M_{\rm rm} = 6 \times
10^7$~\Msun, then $\Mdot \la 4.2 \times 10^{-3}$~\Msun~yr$^{-1}$ and
$\Mdot c^2/2 = 1.2 \times 10^{44}$~erg~s$^{-1}$.  In this case, the
X-ray luminosity exceeds the limit imposed by global energy balance
for all observations except epoch~9.

If the Comptonizing plasma is an ADAF, then the radial velocity of the
accretion flow in this region is $v_r \sim \alpha v_\phi$,
substantially higher than that of the surrounding thin disk (Narayan
\& Yi 1995).  From mass conservation and our estimates of the
accretion rate, we can determine a characteristic Thomson depth,
$\tau_{\rm c}$, for the central region.  Mass conservation in an
accretion disk is given by
\begin{eqnarray}
\Mdot &=& 2\pi r v_r \Sigma \\
      &\sim & 4\pi r^2 \rho v_r,
\label{mass_continuity}
\end{eqnarray}
where $\Sigma = 2h\rho$ is the disk surface density, $h$ is the
half-thickness, and we have set $h \sim r$ for the central region
which we assume to be dominated by ions at the virial temperature.
The density of the central region is roughly
\begin{equation}
\rho \sim \frac{\tau_{\rm c}}{r_s\kappa_e},
\label{ADAF_density}
\end{equation}
where $\kappa_e = 0.4$~cm$^2$g$^{-1}$ is the electron scattering
opacity.  Solving equations~(\ref{mass_continuity})
and~(\ref{ADAF_density}) for the Thomson depth and inserting nominal
values, we obtain
\begin{eqnarray}
\tau_{\rm c} &\sim& 0.4 \pfrac{\alpha}{0.1}^{-1}
                    \pfrac{r_s}{2\times 10^{14}\,{\rm cm}}^{-1/2} \nonumber\\
              & &\times \pfrac{M}{2\times 10^7\,{\rm M}_\odot}^{-3/2}
                  \pfrac{\MMd}{2.5\times 10^5\,{\rm M}_\odot^2{\rm~yr}^{-1}}.
\end{eqnarray}
Despite the crudeness of this estimate, it is encouraging that
$\tau_{\rm c}$ lies within the range of Thomson depths yielded by our
model (Tables~\ref{Ginga_fits}, \ref{RXTE_fits},
and~\ref{new_RXTE_fits}).

A potentially serious problem for this model is the large change in
Thomson depth that we find between epochs~2.1 and~2.2 of the 1998
observations for both sets of response matrices
(Tables~\ref{RXTE_fits} and~\ref{new_RXTE_fits}).  Since changes in
$\rmin$ must occur on time scales that are at least as long as the
thermal time scale and since epochs 2.1--2.4 are fairly closely spaced
over a 240~ks period (Chiang et al.\ 2000; see also
Table~\ref{RXTE_data}), we must adopt a single radius for these
epochs.  For the 1999 response matrices and assuming $\rmin = 2.4
\times 10^{14}$~cm, the Thomson depth must change from $\tau = 0.06$
to $\tau = 2.26$ over $\sim 30$~ks.  Even for the 2002 response
matrices, the Thomson depth must change from $\tau = 0.21$ to $0.69$,
assuming $\rmin = 3.2 \times 10^{14}$~cm.  The necessity for these
large changes can be readily understood.  During these epochs, the
$\lambda5100$\AA\ flux was essentially constant
(Table~\ref{RXTE_data}).  As we found for NGC~3516, in order to have a
nearly constant optical flux, the radius of the Comptonizing region
has to maintain an inverse relationship with the X-ray luminosity
(eq.~[\ref{MMdot}]).  Since the X-ray spectral index increased over
this interval, the radius $r_s$ also had to increase in order to
produce a softer spectrum; and therefore $L_x$ had to decrease.
However, the 2--10~keV X-ray flux increased at the same time, and so
the electron temperature had to decrease by a very large amount to
compensate.  For a fixed Thomson depth, the electron temperature after
the transition would result in an X-ray spectrum that would be much
softer than is observed, and so the Thomson depth had to change
substantially as well.  The time scale for the evolution of the
Thomson depth of the central region should be limited by either the
thermal or accretion (i.e., viscous) time scales.  Assuming a virial
temperature for the ions in the central region, these two time scales
are approximately the same, $t_{\rm acc} \sim t_{\rm th} \sim 10^6$~s
(eq.~[\ref{t_th}]).

One possible mitigating factor here is the relatively sparse coverage
of the optical observations (see \S2).  Multiplying the optical fluxes
for epochs~2.1--2.4 in Table~\ref{RXTE_data} by the factors 0.86,
1.14, 1.07, and 0.98, respectively, yields a nearly constant Thomson
depth of $\tau = 0.44$ and electron temperatures $\kT \simeq
169$--184~keV, assuming $\rmin = 2.4\times 10^{14}$~cm.
Unfortunately, a $> 28$\% change in the $\lambda 5100$\AA\ flux over
the $\sim 30$~ks separating epochs 2.1 and 2.2 does not seem to be
supported by the data (Dietrich et al.\ 2001, see their Fig.~4).  Of
course, it is the UV photons produced near the innermost radius of the
thin disk that are most important for determining the energy balance
of the Comptonizing plasma, and they may not be well correlated with
the emission at longer wavelengths on these time scales.  The disk
flux at $\lambda 5100$\AA\ is produced at a characteristic radius of
$r_{5100} \sim \rmin[T_0/(0.29~{\rm cm~K}/5.1\times 10^{-5}~{\rm
cm})]^{4/3} \sim 3\times 10^{15}$~cm.  Therefore, the optical emission
should be smeared out and delayed relative to the UV emission by $\sim
10^5$~s.

\section{Conclusions}

We have fit simultaneous optical, UV, and X-ray data of NGC 5548 with
reasonable success over a wide range of continuum flux levels using
just a crude model consisting of a stationary, truncated accretion
disk and a spherical hot inner plasma.  Under fairly general
assumptions concerning accretion efficiency, energy balance, and
self-consistency, we are able to place upper limits on the black hole
mass, $M\la 2 \times 10^7$~\Msun\ and the accretion rate, $\Mdot \la
2.5\times 10^5$~M$_\odot^2$yr$^{-1}/M$.  The upper limit on the black
hole mass is less than that derived from reverberation mapping, which
may itself be subject to some systematic uncertainty.  Increasing the
accretion efficiency, perhaps by extracting black hole spin energy,
would relax our constraint on the black hole mass.  We also find that
the disk truncation radius is of order the size of the Comptonizing
plasma and is $\sim 2$--$5 \times 10^{14}$~cm.  This is remarkably
consistent with the size scale inferred in a completely different
manner from the measured EUV to X-ray time lags (Chiang et al.\ 2000).

Although the present model does not provide a perfect description of
the multiwavelength variability of NGC 5548, these calculations do
show the potential for using multiwavelength data for constraining
properties of the disk/black hole system in these objects.  Wider
spectral coverage with more sensitive X-ray observatories will help
enormously in testing models such as this.  In particular, the ability
to measure the thermal Comptonization roll-over at energies $\ga
50$~keV will allow us to abandon our ad hoc temperature limits and the
assumption of a global energy balance condition.  Forthcoming
\INTEGRAL\ observations are the best near-term prospects for better
measuring the thermal Comptonization roll-over in Seyfert 1 galaxies;
and some of the stronger sources --- NGC 4151, IC 4329A, and
MCG$-$6-30-15 --- are among its highest priority targets.  However,
even \INTEGRAL\ does not have the sensitivity to measure changes in
the roll-over energy on the required time scales of $\sim 30$~ks.  For
this, an instrument such as the hard X-ray detector aboard {\em
ASTRO-E2} is required.

\Newton\ observations of NGC~5548 alone may be able to provide
stringent tests of this model.  If the iron K$\alpha$ line is
relativistically broadened, then the EPIC detectors aboard \Newton\
should be able to determine both the disk inclination $i$ and the disk
inner radius, $\rmin$, in units of $GM/c^2$.  Combined with optical/UV
spectroscopy available from the \Newton\ Optical Monitor that can
determine $\rmin$ directly, such observations should be able to
constrain the black hole mass.  However, if $\rmin \gg GM/c^2$, as
suggested by our analysis of the NGC 5548 data, then the iron line
will not be relativistically broadened, and \Newton\ observations will
provide at best a weak upper limit on $M$.

We wish to emphasize again that our model assumes that the disk
itself is stationary, and all optical/UV variability is a result of
the variable incident X-ray flux and varying $\rmin$.  As we noted in
equation (\ref{t_th}), the disk can be out of thermal equilibrium in
the innermost regions on time scales comparable to the observed
variations.  If thermal instabilities are responsible for the
variations in $\rmin$, then our model may not be fully consistent.
Even in the optically emitting regions, the thermal time scale is only
$\sim 2$~yr~$(\alpha/0.1)^{-1}$ for a black hole mass of
$10^7$~{\Msun}.  It may therefore be interesting to develop our model
further to incorporate a non-stationary disk structure.

\acknowledgments 

We thank Matthias Dietrich and Mike Eracleous for providing optical
data from ground-based observations of NGC~5548, and we are grateful
to Pawel Magdziarz for making available the re-extracted \Ginga\ data.
We thank the referee, Andrzej Zdziarski, for helpful comments and
criticisms of the manuscript and for pointing out the changes in the
calculation of the \RXTE/\PCA\ response matrices.  We thank Jim Dove
for providing calculations from his thermal Comptonization Monte Carlo
code; and we thank Jim and Mike Nowak for helpful conversations on the
applicability of these calculations to AGN observations.  We
acknowledge early work with Andrew Frey to model the effect of a
changing disk truncation radius on the optical/UV variability of
NGC~5548.  J.C. was partially supported by NASA ATP grant NAG 5-7723.
O.B. acknowledges support from NASA ATP grant NAG-7075. This research
has made use of data obtained from the High Energy Astrophysics
Science Archive Research Center (HEASARC), provided by NASA's Goddard
Space Flight Center; and it has also made use of the NASA/IPAC
Extragalactic Database (NED) which is operated by the Jet Propulsion
Laboratory, California Institute of Technology, under contract with
NASA.

\appendix

\section{Description of the Thermal Comptonization Monte Carlo}

Seed photons from the disk are drawn assuming that the disk radial
temperature profile is set by contributions from local viscous
dissipation and reprocessing of X-ray flux from the hot plasma.  The
intensity of the emission is assumed to be isotropic at each point on
the surface of the disk, and the energy distribution of the photons is
modified from black body to account for the temperature dependent
empirical color correction we have applied (see \S3 and Paper~II).
For disk photons which intercept the central Comptonizing region, the
optical depth is computed using the full Klein-Nishina cross-section
and assuming a single-temperature relativistic Maxwellian distribution
for the electrons (Poutanen \& Svensson 1996).  A scattering location
along the photon path is drawn from the appropriate distribution; and
if the photon is scattered, then its subsequent trajectory and energy
are drawn from the scalar Compton redistribution function (Poutanen \&
Svensson 1996; Jones 1968).  The optical depth to the boundary of the
Comptonizing region is computed and a new scattering location is
drawn.  This continues until the photon escapes the Comptonizing
region.

If the escaping photon intercepts the disk, then a separate Compton
reflection Monte Carlo is employed.  In this calculation, the incident
photon strikes a semi-infinite slab of material at the appropriate
angle and is scattered by the cold electrons in the slab until it
either re-emerges or is absorbed.  For the material composing the
slab, we adopt the absorption cross-sections and abundances of
Ba{\l}uci\'nska-Church \& McCammon (1992), and assume that the
hydrogen and helium atoms are fully ionized and the remaining elements
are neutral.  This part of the code is functionally identical to the
Monte Carlo calculations used by Magdziarz \& Zdziarski (1995) to
derive parameterizations of the Compton reflection Green functions.
We have tested our code against their expressions and find consistent
reflection spectra.  Photons that escape the Comptonizing region
without hitting the disk and those that emerge from the disk without
re-entering the plasma are included in the final spectra according to
their energy and direction.

The thermal Comptonization part of our Monte Carlo code has been
checked against a two-dimensional iterative scattering (IS) code in
which the integrals for the source functions at each scattering order
are computed numerically.  This IS code has in turn been checked
against the one-dimensional IS code of Poutanen \& Svensson (1996).
We confirm that our Monte Carlo calculation agrees at each scattering
order with the IS code.  Using the Monte Carlo code rather than the IS
code presents several advantages.  The principal one is that it runs
more quickly than the IS code, within a certain accuracy, allowing us
to explore parameter space more efficiently.  Furthermore, interfacing
it with the Compton reflection part of the code is easier since both
methods deal with individual photons.  Lastly, should we decide to
generalize this calculation, it would be a straight-forward matter to
modify the geometry of the Comptonizing region or take into account
curved spacetimes in computing the trajectories of the photons.

In order to find the equilibrium temperature of the Comptonizing
electrons, we proceeded iteratively, using the difference between the
model and observed fluxes at a nominal wavelength (in this case,
$\lambda 5100$\AA) as the main convergence criterion.  For the
calculation of an individual spectrum, the radial Thomson depth
$\tau_{\rm r}$, the radius of the Comptonizing region $r_s$, the disk
inner radius $\rmin$, the central black hole mass, the accretion rate,
and initial estimates of the electron temperature and X-ray luminosity
are taken from our semi-analytic calculations of \S3.  For the first
iteration, we assume a nominal value for the disk albedo, and we use
the expressions from Zdziarski et al.\ (1999), modified to include the
contribution from local viscous dissipation, to determine the
effective temperature as a function of disk radius.  On subsequent
iterations, the model and observed optical fluxes are compared, and
the electron temperature (and therefore the X-ray luminosity) is
adjusted.  Since we are computing the Compton reflection spectrum in
the disk at each iteration, we obtain the disk albedo as a function of
radius which we use as input for the subsequent iteration.
Convergence has been achieved when both the model optical flux agrees
with the observed value and the X-ray flux has converged to within a
specified tolerance---typically 5\% for each.  As with our
semi-analytic calculations, our Monte Carlo calculations do not
consider pair balance or the presence of a non-thermal component in
determining the properties of the Comptonizing plasma.

\clearpage

\clearpage

\begin{table}
\caption{Optical and UV continuum fluxes and X-ray continuum
parameters from the 1989--90 ground-based/\IUE/\Ginga\ data.
\label{IUE_Ginga_data}}
\bigskip
\begin{tabular}{ccccccccc}
\hline\hline
epoch & JD$-2440000$ & $F_{5100}$\tablenotemark{a} 
      & $F_{2670}$\tablenotemark{a} & $F_{1840}$\tablenotemark{a} 
      & $F_{1350}$\tablenotemark{a} & $F_{210}$\tablenotemark{b} & 
      $\Gamma$\tablenotemark{c} & $R$\tablenotemark{c,d}\\
\hline
1 & 7535--7539 & 0.78 $\pm$ 0.08 &  2.22 $\pm$ 0.19 &  4.52 $\pm$ 0.22 
  &  7.45 $\pm$ 0.33 & 3.39 & 1.56 $\pm$ 0.19 & $0.32^{+0.72}_{-0.31}$\\
2 & 7556       & 0.62 $\pm$ 0.06 &  2.09 $\pm$ 0.20 &  3.91 $\pm$ 0.22 
  &  6.18 $\pm$ 0.33 & 5.43 & 1.72 $\pm$ 0.03 & $0.58 \pm 0.20$ \\
3 & 7685       & 0.67 $\pm$ 0.06 &  1.72 $\pm$ 0.23 &  2.88 $\pm$ 0.31 
  &  3.94 $\pm$ 0.46 & 4.71 & 1.71 $\pm$ 0.10 & $0.44^{+0.41}_{-0.31}$\\
4 & 8036--8037 & 0.50 $\pm$ 0.06 &  1.31 $\pm$ 0.36 &  2.56 $\pm$ 0.30 
  &  3.84 $\pm$ 0.35 & 3.87 & 1.68 $\pm$ 0.06 & $0.28^{+0.24}_{-0.12}$\\
5 & 8039--8041 & 0.51 $\pm$ 0.07 &  1.44 $\pm$ 0.25 &  2.85 $\pm$ 0.36 
  &  4.25 $\pm$ 0.42 & 3.53 & 1.78 $\pm$ 0.06 & $0.97^{+0.52}_{-0.42}$\\
6 & 8047       & 0.49 $\pm$ 0.07 &  1.40 $\pm$ 0.19 &  2.59 $\pm$ 0.30 
  &  4.21 $\pm$ 0.35 & 2.94 & 1.75 $\pm$ 0.08 & $0.97^{+0.72}_{-0.39}$\\
7 & 8056       & 0.50 $\pm$ 0.06 &  1.53 $\pm$ 0.17 &  2.80 $\pm$ 0.21 
  &  4.35 $\pm$ 0.24 & 4.95 & 1.74 $\pm$ 0.04 & $0.38^{+0.24}_{-0.20}$\\
8 & 8068       & 0.39 $\pm$ 0.05 &  1.10 $\pm$ 0.13 &  1.64 $\pm$ 0.30 
  &  2.23 $\pm$ 0.35 & 2.44 & 1.53 $\pm$ 0.04 & $< 0.27$\\
9 & 8077       & 0.34 $\pm$ 0.05 &  0.96 $\pm$ 0.20 &  1.30 $\pm$ 0.30 
  &  1.74 $\pm$ 0.35 & 1.93 & 1.71 $\pm$ 0.14 & $0.87^{+0.99}_{-0.60}$\\
\hline
\end{tabular}
\tablenotetext{a}{Units are
$10^{-14}$~erg~cm$^{-2}$s$^{-1}$\AA$^{-1}$.  The flux uncertainties
are 1-$\sigma$.}  \tablenotetext{b}{Units are
$10^{-11}$~erg~cm$^{-2}$s$^{-1}$.  Throughout this paper, the 2--10
keV flux is computed for the {\em underlying} X-ray continuum which is
assumed to be a cut-off power-law in the {\sc pexrav} model.
Therefore, these fluxes exclude the Fe K$\alpha$ emission, the Compton
reflection component, and corrects for the effects of absorption.}
\tablenotetext{c}{Uncertainties are computed for $\Delta \chi^2 =
2.706$.}  \tablenotetext{d}{Compton reflection fraction.}
\end{table}

\begin{table}
\caption{Model parameters fit to the 1989--90
ground-based/\IUE/\Ginga\ data assuming $M = 1\times 10^7$~\Msun,
$\Mdot = 0.02$~\Msun~yr$^{-1}$, and $\cos i=0.7$. Note that $\rmin$ and
$T_0$ depend only on $\cos i$.
\label{Ginga_fits}}
\begin{tabular}{cccccccc}
\hline\hline
epoch & $\rmin$\tablenotemark{a} & $T_0$ ($10^4$~K) 
      & $r_s$\tablenotemark{a} 
      & $\kT$ (keV) & $\tau$ 
      & $L_x$\tablenotemark{b}
      & $L_{\rm tr}$\tablenotemark{b}\\
\hline
1 & 2.20 & 4.0 & 2.07 &  683 & 0.13 & 4.65 & 0.11  \\
2 & 1.85 & 4.2 & 2.07 &  153 & 0.72 & 3.03 & 0.13  \\
3 & 3.36 & 2.8 & 3.70 &   66 & 1.64 & 2.10 & 0.07  \\
4 & 2.31 & 3.3 & 2.42 &   76 & 1.54 & 1.84 & 0.10  \\
5 & 2.07 & 3.6 & 2.47 &   96 & 1.05 & 1.68 & 0.11  \\
6 & 1.97 & 3.7 & 2.22 &  226 & 0.44 & 1.77 & 0.12  \\
7 & 1.96 & 3.7 & 2.19 &   37 & 2.46 & 1.90 & 0.12  \\
8 & 2.57 & 2.8 & 2.17 &   85 & 1.82 & 1.44 & 0.09  \\
9 & 2.62 & 2.7 & 2.76 &   50 & 2.05 & 0.80 & 0.09 \\
\hline
\end{tabular}
\tablenotetext{a}{Units are $10^{14}$~cm.}
\tablenotetext{b}{Units are $10^{44}$~erg~s$^{-1}$.}
\end{table}

\clearpage

\begin{table}
\caption{Optical continuum fluxes and X-ray continuum parameters from 
the 1998 ground-based/RXTE data.
\label{RXTE_data}}
\bigskip
\begin{tabular}{ccccccccc}
\hline\hline
epoch & JD$-2450000$ & $F_{5100}$\tablenotemark{a} &
      \multicolumn{3}{c}{1999 PCA responses\tablenotemark{b}} & 
      \multicolumn{3}{c}{2002 PCA responses\tablenotemark{c}} \\
 & & & $F_{210}$\tablenotemark{d} & $\Gamma$\tablenotemark{e} & $R$\tablenotemark{e}
     & $F_{210}$\tablenotemark{d} & $\Gamma$\tablenotemark{e} & $R$\tablenotemark{e}\\
\hline
1.0 & 979.96--980.40 & $1.16 \pm 0.06$ & 8.0 & $1.86 \pm 0.03$ 
    & $0.34_{-0.18}^{+0.18}$ & 7.9 & $1.72 \pm 0.03$ & $0.20_{-0.13}^{+0.17}$ \\
2.1 & 985.09--985.40 & $1.17 \pm 0.09$ & 7.8 & $1.86 \pm 0.03$ 
    & $0.38_{-0.17}^{+0.19}$ & 7.7 & $1.72 \pm 0.03$ & $0.24_{-0.14}^{+0.17}$ \\
2.2 & 985.69--986.41 & $1.18 \pm 0.09$ & 9.8 & $1.93 \pm 0.02$ 
    & $0.46 \pm 0.12$ & 9.5 &    $1.80 \pm 0.02$   & $0.33_{-0.10}^{+0.11}$ \\
2.3 & 986.65--987.40 & $1.16 \pm 0.09$ & 9.5 & $1.90_{-0.01}^{+0.02}$ 
    & $0.37_{-0.04}^{+0.10}$ & 9.3 & $1.77_{-0.03}^{+0.02}$ 
    & $0.25_{-0.10}^{+0.11}$ \\
2.4 & 987.63--987.86 & $1.24 \pm 0.09$ & 9.1 & $1.91_{-0.03}^{+0.04}$ 
    & $0.46_{-0.13}^{+0.18}$ & 8.8 & $1.78_{-0.04}^{+0.05}$ 
    & $0.34_{-0.20}^{+0.28}$ \\
3.1 & 994.36--994.89 & $1.03 \pm 0.04$ & 6.9 & $1.80_{-0.04}^{+0.05}$ 
    & $0.29_{-0.14}^{+0.29}$ & 6.8 & $1.66_{-0.04}^{+0.05}$ 
    & $0.17_{-0.17}^{+0.24}$ \\ 
3.2 & 995.64--996.13 & $1.01 \pm 0.08$ & 5.6 & $1.80_{-0.02}^{+0.03}$ 
    & $0.38_{-0.09}^{+0.18}$ & 5.5 & $1.65_{-0.03}^{+0.04}$ 
    & $0.23_{-0.14}^{+0.17}$ \\
\hline
\end{tabular}

\tablenotetext{a}{Units are $10^{-14}$~erg~cm$^{-2}$s$^{-1}$\AA$^{-1}$;
1-$\sigma$ uncertainties.}
\tablenotetext{b}{Using {\sc pcarsp v2.36} and {\sc pcarmf v3.5}.}
\tablenotetext{c}{Using {\sc pcarsp v8.0} and {\sc pcarmf v8.0}.}
\tablenotetext{d}{Units are $10^{-11}$~erg~cm$^2$s$^{-1}$.}
\tablenotetext{e}{Uncertainties are computed for $\Delta \chi^2 = 2.706$.}

\end{table}

\begin{table}
\caption{Model parameters fit to the 1998 ground-based/RXTE data for
$\rmin = 2.4 \times 10^{14}$~cm and $\rmin = 3.2\times 10^{14}$~cm (in
parentheses).  These parameters were derived from spectral fits that
used PCA response matrices generated by {\sc pcarsp v2.36} and {\sc
pcarmf v3.5} (Chiang et al.\ 2000).
\label{RXTE_fits}}
\bigskip
\begin{tabular}{ccccccc}
\hline\hline
epoch & $r_s$ ($10^{14}$~cm) & $\kT$ (keV) & $\tau$ 
      & $L_x$\tablenotemark{a} \\
\hline
1.0 & 3.78 (5.02) &  554 (136) & 0.09 (0.63) & 5.15 (4.04) \\
2.1 & 3.79 (5.02) &  684 (168) & 0.06 (0.49) & 5.21 (4.09) \\
2.2 & 4.64 (5.99) &   27 ( 12) & 2.26 (3.80) & 4.28 (3.94) \\
2.3 & 4.24 (5.52) &   89 ( 12) & 0.90 (3.99) & 4.57 (3.75) \\
2.4 & 4.38 (5.81) &  204 ( 27) & 0.36 (2.31) & 4.92 (3.86) \\
3.1 & 3.24 (4.31) &  676 (229) & 0.07 (0.37) & 5.00 (3.95) \\
3.2 & 3.24 (4.31) & 1552 (526) & 0.02 (0.11) & 4.87 (3.84) \\
\hline
\end{tabular}
\tablenotetext{a}{Units are $10^{44}$~erg~s$^{-1}$.}
\end{table}

\begin{table}
\caption{Model parameters fit to the 1998 ground-based/RXTE data for
$\rmin = 3.2\times 10^{14}$~cm. These parameters were derived from
spectral fits that used PCA response matrices generated by {\sc pcarsp
v8.0} and {\sc pcarmf v8.0}.
\label{new_RXTE_fits}}
\bigskip
\begin{tabular}{ccccccc}
\hline\hline
epoch & $r_s$ ($10^{14}$~cm) & $\kT$ (keV) & $\tau$ 
      & $L_x$\tablenotemark{a} \\
\hline
1.0 & 3.72 & 346 & 0.25 & 5.57\\
2.1 & 3.72 & 393 & 0.21 & 5.63\\
2.2 & 4.34 & 135 & 0.69 & 4.85\\
2.3 & 4.07 & 164 & 0.59 & 5.05\\
2.4 & 4.16 & 305 & 0.26 & 5.48\\
3.1 & 3.43 & 285 & 0.38 & 5.05\\
3.2 & 3.39 & 471 & 0.19 & 4.98\\
\hline
\end{tabular}
\tablenotetext{a}{Units are $10^{44}$~erg~s$^{-1}$.}
\end{table}


\clearpage

\begin{thebibliography}{}
\bibitem{B92} Ba{\l}uci\'nska-Church, M., \& McCammon, D. 1992, ApJ,
   400, 699
\bibitem{B99} Beloborodov, A. M. 1999, in High Energy Processes in
   Accreting Black Holes, eds. J. Poutanen and R. Svensson, (San
   Francisco: Astronomical Society of the Pacific), 295
\bibitem{C00} Chiang, J., et al.\ 2000, ApJ, 528, 292
\bibitem{C01} Chiang, J., \& Blaes, O.\ 2001, ApJ, 557, L15 (Paper~I)
\bibitem{CB02} Chiang, J., \& Blaes, O.\ 2002, in Proc. JHU/LHEA 
   Workshop on X-ray Emission from Accretion onto Black Holes, 
   ed. T. Yaqoob and J. H. Krolik, published electronically at
   {\tt http://www.pha.jhu.edu/groups/astro/workshop2001} 
   {\tt /papers/chiang\_j.ps}
\bibitem{C02} Chiang, J.\ 2002, ApJ, 572, 79 (Paper~II)
\bibitem{C91} Clavel, J., et al.\ 1991, ApJ, 366, 64
\bibitem{C92} Clavel, J., et al.\ 1992, ApJ, 393, 113
\bibitem{C98} Collier, S., et al.\ 1998, ApJ, 500, 162
\bibitem{CS91} Collin-Souffrin, S.\ 1991, A\&A, 249, 344
\bibitem{C99} Coppi, P. S.\ 1999, in High Energy Processes in
   Accreting Black Holes, eds. J. Poutanen and R. Svensson, (San
   Francisco: Astronomical Society of the Pacific), 375
\bibitem{CC91} Courvoisier, T. J.-L., \& Clavel, J. 1991, A\&A, 248, 389
\bibitem{D01} Dietrich, M., et al.\ A\&A, 371, 79
\bibitem{D92} Done, C., Mulchaey, J.~S., Mushotzky, R.~F., 
   \& Arnaud, K.~A.\ 1992, ApJ, 395, 275
\bibitem{D97} Dove, J. B., Wilms, J., \& Begelman, M. C., 1997, ApJ,
   487, 747
\bibitem{E00} Edelson, R. E., et al. 2000, ApJ, 534, 180
\bibitem{G99} Gammie, C. F.\ 1999, ApJ, 522, L57
\bibitem{G94} Guainazzi, M., Matsuoka, M., Piro, L., Mihara, T., \&
   Yamauchi, M., 1994, ApJ, 436, L35
\bibitem{H01} Hubeny, I., Blaes, O., Krolik, J. H., Agol, E.\ 2001, ApJ, 
   559, 680
\bibitem{J68} Jones, F. C.\ 1968, Phys. Rev., 167, 1159
\bibitem{K91} Krolik, J. H., et al.\ 1991, ApJ, 371, 541
\bibitem{K99} Krolik, J. H.\ 1999, ApJ, 515, L73
\bibitem{K01} Krolik, J. H.\ 2001, ApJ, 551, 72
\bibitem{M95} Magdziarz, P., \& Zdziarski, A. A.\ 1995, MNRAS, 273, 837
\bibitem{M98} Magdziarz, P., et al.\ 1998, MNRAS, 301, 179
\bibitem{M01} Malzac, J., Beloborodov, A. M., \& Poutanen, J.\ 2001, 
   MNRAS, 326, 417
\bibitem{Mu95} Mushotzky, R. F., et al.\ 1995, MNRAS, 272, L9
\bibitem{N91} Nandra, K., et al.\ 1991, MNRAS, 248, 760
\bibitem{N98} Nandra, K., et al.\ 1998, ApJ, 505, 594
\bibitem{N00} Nandra, K., et al.\ 2000, ApJ, 544, 734
\bibitem{N95} Narayan, R., \& Yi, I. 1995, ApJ, 452, 710
\bibitem{P91} Peterson, B. M., et al.\ 1991, ApJ, 368, 119
\bibitem{P00} Peterson, B. M., \& Wandel, A. 2000, ApJ, 540, L13
\bibitem{P96} Poutanen, J., \& Svensson, R.\ 1996, ApJ, 470, 249
\bibitem{P97} Poutanen, J., Krolik, J. H., \& Ryde, F.\ 1997,
   MNRAS, 292, L21
\bibitem{S76} Shapiro, S. L., Lightman, A. P., \& Eardley, D. N.\ 1976, 
   ApJ, 204, 187
\bibitem{S95} Skibo, J. G., Dermer, C. D.\ 1995, ApJ, 455, L25
\bibitem{W90} Wamsteker, W., et al.\ 1990, ApJ, 354, 446
\bibitem{W01} Wilms, J., et al.\ 2001, MNRAS, 328, L27
\bibitem{Z99} Zdziarski, A. A., Lubi\'nski, P., \& Smith, D. A., 1999,
   MNRAS, 303, L11
\bibitem{Z00} Zdziarski, A. A., Poutanen, J., \& Johnson, W. N., 2000,
   ApJ, 542, 703
\end{thebibliography}
\end{document}